\documentclass{article}
\pdfoutput=1 

\usepackage{geometry}
\usepackage[utf8]{inputenc}
\usepackage[numbers,sort]{natbib}
\usepackage{fancybox}
\usepackage{tabularx}
\usepackage{booktabs}
\usepackage{graphicx}
\usepackage{adjustbox}
\usepackage{rotating}
\usepackage[noend]{algpseudocode}
\usepackage{amsmath}
\usepackage{amsthm}
\usepackage{amsfonts}
\usepackage{mathrsfs}
\usepackage{tikz}
\usepackage{listings}
\usepackage{appendix}
\usepackage{multicol}
\usepackage{ragged2e}
\usepackage{setspace}

\usepackage{stmaryrd} 
\usepackage{amssymb} 
\usepackage{url}
\usepackage{color}
\usepackage{float}
\floatstyle{plain} 
\newfloat{Code}{H}{myc}
\let\origthelstnumber\thelstnumber
\makeatletter
\newcommand*\Suppressnumber{%
  \lst@AddToHook{OnNewLine}{%
    \let\thelstnumber\relax%
     \advance\c@lstnumber-\@ne\relax%
    }%
}

\newcommand*\Reactivatenumber{%
  \lst@AddToHook{OnNewLine}{%
   \let\thelstnumber\origthelstnumber%
   \advance\c@lstnumber\@ne\relax}%
}
 
\definecolor{codegreen}{rgb}{0,0.6,0}
\definecolor{codegray}{rgb}{0.5,0.5,0.5}
\definecolor{codepurple}{rgb}{0.58,0,0.82}
\definecolor{backcolour}{rgb}{0.97,0.97,0.97}
 
\lstdefinestyle{mystyle}{
    backgroundcolor=\color{backcolour},   
    commentstyle=\color{codegreen},
    keywordstyle=\color{magenta},
    numberstyle=\tiny\color{codegray},
    stringstyle=\color{codepurple},
    basicstyle=\ttfamily\scriptsize,
    breakatwhitespace=false,         
    breaklines=true,                 
    captionpos=b,                    
    keepspaces=true,                 
    numbers=left,                    
    numbersep=5pt,                  
    showspaces=false,                
    showstringspaces=false,
    showtabs=false,                  
    tabsize=2
}
 
\usepackage{adjustbox}
\usepackage{stmaryrd} 
\usetikzlibrary{shapes,snakes, positioning, arrows, calc}
\pgfarrowsdeclarecombine{twotriang}{twotriang}%
{triangle 45}{triangle 45}{triangle 45}{triangle 45}
\pgfarrowsdeclarecombine{twotriangR}{twotriangR}%
{triangle 45 reversed}{triangle 45 reversed}{triangle 45 reversed}{triangle 45 reversed}
\pgfarrowsdeclarecombine{combtriang}{combtriang}%
{twotriangR}{twotriang}{twotriangR}{twotriang}
\theoremstyle{definition}

\usepackage[colorinlistoftodos]{todonotes}

\title{MetaChem: An Algebraic Framework for Artificial Chemistries}
\author{Penelope Faulkner Rainford\\
Department of Chemistry, University of York, UK;\\
York Cross-disciplinary Centre for Systems Analysis, University of York, UK;\\
Business School, University of Hull, UK.\\
p.rainford@hull.ac.uk\\~\\
Angelika Sebald\\
Department of Chemistry, University of York, UK;\\
York Cross-disciplinary Centre for Systems Analysis, University of York, UK.\\
angelika.sebald@york.ac.uk\\~\\
Susan Stepney \\
Department of Computer Science, University of York, UK;\\
York Cross-disciplinary Centre for Systems Analysis, University of York, UK.\\
susan.stepney@york.ac.uk\\
}
\date{}

\begin{document}
\maketitle

\begin{abstract}
We introduce MetaChem, a language for representing and implementing Artificial Chemistries. We motivate the need for modularisation and standardisation in representation of artificial chemistries. We describe a mathematical formalism for Static Graph MetaChem, a static graph based system. MetaChem supports different levels of description, and has a formal description; we illustrate these using StringCatChem, a toy artificial chemistry. We describe two existing Artificial Chemistries -- Jordan Algebra AChem and Swarm Chemistries -- in MetaChem, and  demonstrate how they can be combined in several different configurations by using a MetaChem environmental link.   MetaChem provides a route to standardisation, reuse, and composition of Artificial Chemistries and their tools.\bigskip

\end{abstract}


\section{Introduction}

The field of Artifical Chemistry covers many rich and diverse systems \citep{Banzhaf-2015-AChem}.
Yet practitioners can struggle to talk about specific systems in the context of the whole field. It is hard to make comparisons between even intuitively quite similar systems. 
For example, consider a toy AChem that concatenates strings of letters from the Roman alphabet; now consider another toy AChem that combines lists of integers with values between 1 and 26.  There is a clear mapping between the two sets of particles, and they use equivalent linking rules. However, such similarities can be hidden, for example by describing one as a string system and the other as a list system. Then, despite the underlying equivalence of the particles and linking rules, their overall behaviour might be very different if, say, strings can decompose but the lists cannot, or if, say, lists move in simulated space and use collisions to determine linking, while strings interact in a well mixed tank.
Are they fundamentally equivalent, or fundamentally different, and where do the differences lie?
In other cases, systems may have similar goals but very different components, algorithms and implementations. 
This makes building even a basic classification system challenging.

To tackle this issue, \cite{Dittrich2001-hw} declare that an artificial chemistry (AChem) is described by the triplet $(S,R,A)$
a set of particles $S$, rules for reactions $R$, and the algorithm $A$. 
The particles and reaction rules are reasonably clearly circumscribed,
but all the other aspects of the system are combined in $A$ as part of the algorithm, covering such diverse concepts as spatiality, rule application, environmental conditions, timing, and logging.

Despite its apparent generality, the $(S,R,A)$ model does not accommodate all systems that practitioners may want to regard as AChems.
A different conceptual view of artificial chemistry may be 
more inclusive of all AChem systems. 
Consider: an artificial chemistry is a system with minimal components designed to use or explore the higher order emergent properties of their interactions. This conceptualisation allows for AChems with purely kinetic interactions, and for ones that do not distinguish between the $R$ and $A$ aspects. 
It also acknowledges that some systems are distinguished as different due to the differences in $A$ rather than $S$ or $R$. 

In the $(S,R,A)$ model, many features of more recent chemistries are lumped into $A$. This one component contains huge amounts of important parts of an AChem, bundled as ``everything else". 
Aspects of a system that may indicate intent, or over-design towards a goal, can be lost in $A$.

Additionally, it is not always clear how to partition the design of an AChem into these three components.
If a system has to use an extra reaction to make membranes possible, where do we find this in $(S,R,A)$? 
If a system is probabilistic based on a property of the environment, is that in $R$ for the reaction, or part of $A$ the algorithm? 
If different designers make different decisions for the same feature, it is hard to compare their systems. For example, there are now three different versions of random boolean network-base subsymbolic AChems  \cite{Faulconbridge2009-ch, Krastev2016-kb, Watson2019-oj},
and subversions within each version, 
These all have essentially the same underlying $S$, but subtly different bonding rules $R$, and a variety of very different $A$ used to explore different properties.
Are the different results due to the changes in $R$ or in $A$?
Overall, it is impossible to rigorously define and quantify differences or similarities between systems with such a crude language of only three words.

The $(S,R,A)$ description was a good tool when it was developed, when AChems were often still small toy systems, or even just thought experiments. It was  sufficient for the many early ``proof of concept" AChems that demonstrate that an artificial system can produce cell-like objects capable of self-maintenance and self-replication. However, as AChems move into new realms of being used for computation and as the basis for open-ended evolution systems, we need to be able to analyse them more rigorously, and make comparisons between alternative models, and reuse components between models. We need to develop a new, more sophisticated descriptive framework for artificial chemistries.

Here we present MetaChem, a formal description language for AChems, which allows us to model, standardise, compare, and combine, diverse AChem systems.
The structure of the paper is as follows.
In section~\ref{sec:properties} we discuss the objects and actions common to AChems, to motivate our design.
In section~\ref{sec:module} we introduce the basic graph structure of MetaChem, with the various node and edge types that can be used to construct an AChem definition.
In section~\ref{sec:overview} we demonstrate how this graph structure can be used to define an AChem at different levels of description.
In section~\ref{sec:static} we describe the internal structure of the executable nodes, showing how the graph forms a program to execute an AChem;
a mathematical definition is provided in appendix~\ref{sec:maths}.
In sections~\ref{sec:jaachem} and~\ref{sec:swarm} we recast two widely differing AChems from the literature into MetaChem, to demonstrate its breadth of applicability.
In section~\ref{sec:nested} we combine these two widely AChems to produce a family of nested AChems, to demonstrate to combinatorial power of MetaChem.
We conclude with a discussion of future extensions to MetaChem to allow dynamically changing graphs.

\section{Properties of Artificial Chemistries}\label{sec:properties}
There are axiomatic concepts that we build on in the field of artificial chemistries. AChems start with  small components interacting to generate our systems. Analysis tends to focus on the emergent properties and behaviours of these systems. To differentiate an AChem from an Individual Based Model \citep{Grimm-2005-IBM}, we add requirements for simplicity (easy to describe) and tractability (easy to compute) in our particles and their interactions. The intention is that these systems work over large collections of individuals and over long time periods, although most are currently limited by computational capability.

From these concepts we identify many common elements of AChem systems, and use these as the basis for a bottom-up approach to systematic modularisation of AChem systems. 
Individual \textit{particles} and their \textit{interactions} are our primary focus. 
These are present throughout AChem systems. Systems also have other variables, properties and values. Much like in real chemistry, we separate the description of the ``glassware'' from our consideration of its
particle contents. We separate  these other values and properties into an \textit{environment}. We can have multiple \textit{containers} in our systems,
which allow us to isolate particles and move them, analogous
to the ``beakers", ``tubes", and ``valves" comprising the
``glassware", or membranes and compartments in biological cells. 
These components comprise the ``things" in our systems, Table~\ref{tab:parts}.

\begin{table}
\centering
\begin{tabular}{lcc}\toprule
& Primary Focus & Auxiliaries\\
\midrule
Objects & Particles & Variables\\
Containers & Tanks & Environment\\
\bottomrule
\end{tabular}
\caption{\label{tab:parts}Common parts of Artificial Chemistry Systems}
\end{table}


There are also commonalities in the algorithms of AChems
(and often their implementations) that we abstract out in our
framework. 
Control flows, related to time and generations, occur in most systems: some systems update across all objects in the system at once; others continuously update objects at random. 
If we can identify the modularised control that produces these timing systems, designers could switch between them. 
This would then allow designers to focus on the new AChem-specific features of their design, whilst exploiting pre-existing elements to implement less unique aspects of their systems.

We define our control flow in relation to how we divide our ``things". 
We \textit{modify particles}, similar to reactions and interactions in chemistry. 
We \textit{record observations} of our system. 
We  \textit{modify our environment}, such as by changing the temperature of the system. 
We \textit{move} particles around our system. 
We \textit{decide} which of these things we should do next. 
These control flow actions form the building blocks of our MetaChem.

\section{Modularisation: Components of an Artificial Chemistry System\label{sec:module}}

 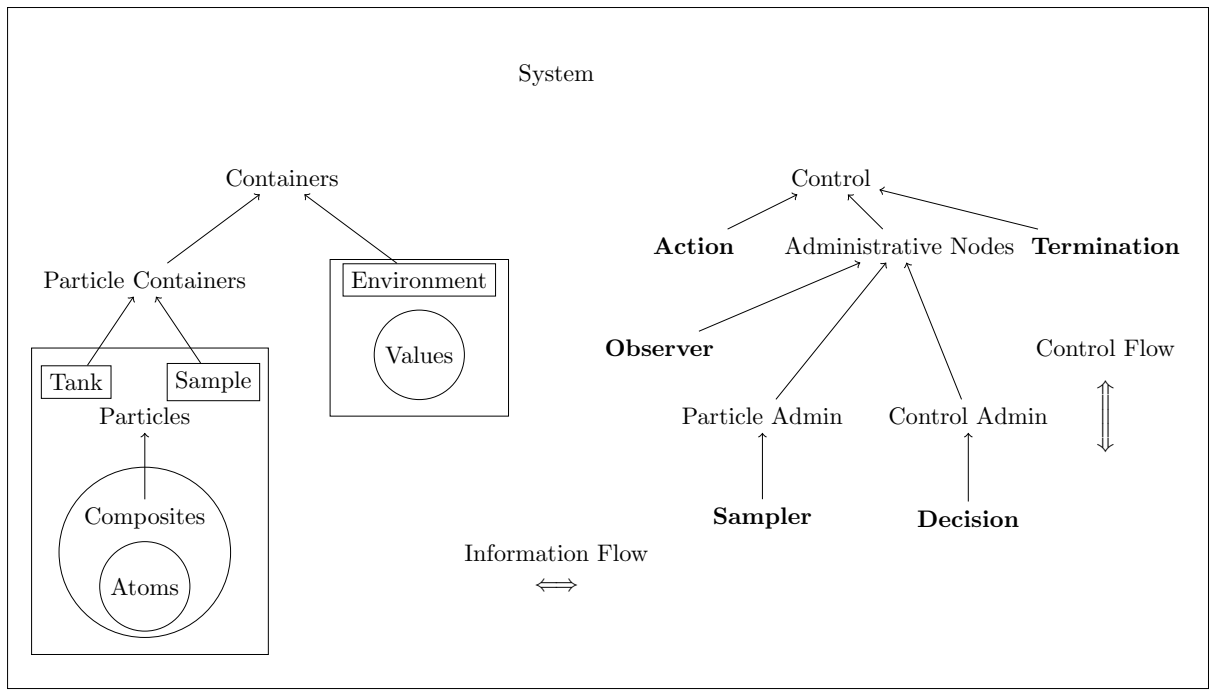
\begin{figure}
\centering
\resizebox{\textwidth}{!}{
\begin{tikzpicture}
\draw (0,0) rectangle (17.5,10);
\node(System) at (8,9) {System};
\node(Box) at (4,7.5) {Containers};
\node(PBox) at (2,6) {Particle Containers};
\node[draw] (Env) at (6,6) {Environment};
\node[draw] (Tank) at (1,4.5) {Tank};
\node[draw] (Sample) at (3,4.5) {Sample};
\node(Part) at (2,4) {Particles};
\node(Comp) at (2,2.5) {Composites};
\node[shape=circle,draw] at (2,1.5) {Atoms};
\node[shape=circle,draw] at (6,4.9) {Values};
\draw (4.7,6.3) rectangle (7.3,4);
\draw (0.35,0.5) rectangle (3.8,5);
\draw (2,2) circle [radius=1.25];
\draw[->] (Tank) -- (PBox);
\draw[->] (Sample) -- (PBox);
\draw[->] (Comp) -- (Part);
\draw[->] (PBox) -- (Box);
\draw[->] (Env) -- (Box);

\node(Control) at (12,7.5) {Control};
\node(Act) at (10,6.5) {\textbf{Action}};
\node(Admin) at (13,6.5) {Administrative Nodes};
\node(Cont) at (14,4) {Control Admin};
\node(Decision) at (14,2.5) {\textbf{Decision}};
\draw[->] (Act) -- (Control);
\node(PAdmin) at (11,4) {Particle Admin};
\node(Sampler) at (11,2.5) {\textbf{Sampler}};
\node(Observer) at (9.5,5) {\textbf{Observer}};
\node(Termination) at (16,6.5) {\textbf{Termination}};
\draw[->] (Termination) -- (Control);
\draw[->] (Admin) -- (Control);
\draw[->] (Decision) -- (Cont);
\draw[->] (Cont) -- (Admin);
\draw[->] (Observer) -- (Admin);
\draw[->] (PAdmin) -- (Admin);
\draw[->] (Sampler) -- (PAdmin);

\node at (8,2) {Information Flow};
\node at (8,1.5) {$\iff$};
\node at (16,5) {Control Flow};
\node at (16,4) {$\Bigg\Updownarrow$};
\end{tikzpicture}}
\caption{Conceptual structure of modularisation of Artificial Chemistries}
\label{fig:structure}
\end{figure}


We arrange these concepts into the structure shown in Figure~\ref{fig:structure}, which we use to build a graph-based formalism. We have the overarching concepts of the System, made up of the elements formalised as graph \textit{nodes} (Containers, Control), and as graph \textit{edges} (Information Flow, Control Flow). 

Control items are static nodes in the graph: their location and connectivity is defined at the start of a ``run'', and remain unchanged as the AChem executes. 
(Future versions will support dynamic connectivity; see \S\ref{sec:dynamic}.)
These control nodes are connected by Control Flow edges, which together define the system's algorithm.


Containers are also static nodes in the graph.
They map to (``contain'') the dynamically changing particles and environmental values in the system. 

 Information Flow edges allows control to influence the connected containers' states (that is, contents). 
 Information can flow in either direction along an edge: \textit{read} or \textit{pulled} from containers' states to the control node, and \textit{pushed} from the control node to update containers' states.

These general concepts are captured by different types of graph nodes (Figure~\ref{tab:legendN}), and types of graph edges (Figure~\ref{tab:legendE}). 
We can use these to define graphs that provide a view of our systems. 
We can provide different views using graphs at different levels:
from a high-level overview, then by expanding nodes down to levels with greater detail
(section~\ref{sec:overview}).

\subsection{Particles}
The most fundamental parts of our systems are the \textit{particles}. These and their emergent properties and behaviours are the focus of our studies. These can usually be split into two subsets: \textit{atoms} and \textit{composites}. 
Some AChems may have only atoms and due to lack of physical bonding rules may not seem to form composites. Others may be symbolic and assume that all particles are complex and that all the symbols represent composites.

\paragraph*{Atomic particles:}the most basic particles; they can not be divided or broken down into smaller parts. Any internal structure of the atoms \citep{Faulkner2018-ex} is indivisible.

\noindent
{\it Examples:} Atoms can take many forms: characters in a string chemistry, instructions in an automata chemistry, or symbols that are not the `one' in a one-to-many symbolic production rule in a symbolic chemistry.


\paragraph*{Composite particles:} these are made of combinations of atoms. In symbolic AChems the atoms making up a composite particle may be hidden or unknown. 

\noindent
{\it Examples:} Composites would be strings in string chemistries, programs in automata chemistries, or symbols that result from many to one rules in symbolic chemistries.

\begin{figure}[t]
\begin{tabular}{cb{0.75\textwidth}}
\toprule
Element & Description\\
\midrule
Containers\\
\midrule
\begin{tikzpicture}
\node[shape=rectangle,draw, label=center:{\textbf{T}}] {\phantom{\textbf{V}}};
\end{tikzpicture} & \textbf{Tank}: particle container\\[1mm]
\begin{tikzpicture}
\node[shape=rectangle,draw, label=center:{\textbf{S}}] {\phantom{\textbf{V}}};
\end{tikzpicture} & \textbf{Sample}: particle container of editable particles\\[1mm]
\begin{tikzpicture}
\node[shape=rectangle,draw, label=center:{\textbf{V}}] {\phantom{\textbf{V}}};
\end{tikzpicture} &  \textbf{Environment}: container of non-particle variables and information in the system.\\
\midrule
Control\\
\midrule
\begin{tikzpicture}
\node[shape=diamond,draw] {\textbf{s}};
\end{tikzpicture} & \textbf{Sampler}: Information administration node that moves particles between containers\\[1mm]
\begin{tikzpicture}
\node[shape=diamond,draw] {\textbf{o}};
\end{tikzpicture} & \textbf{Observer}: Information administration node that observes particles in containers, and saves summary statistics into the environment\\[1mm]
\begin{tikzpicture}
\node[shape=regular polygon,regular polygon sides = 3,draw, label=center:{\textbf{d}}] {\phantom{.}};
\end{tikzpicture}  & \textbf{Decision}: Control administration node  decides on control flow path based on the state of particles and the environment \\[1mm]
\begin{tikzpicture}
\node[shape=circle,draw] {\textbf{a}};
\end{tikzpicture}  & \textbf{Action}: Control node that performs actions on particles based on state of particles and environment \\[1mm]
\begin{tikzpicture}
\node[shape=circle,fill=black,draw] {};
\end{tikzpicture} & \textbf{Termination}: Control node where processing terminates\\
\bottomrule
\end{tabular}
\caption{Graph node types, and their graphical representation, used in MetaChem.
The initial control node (typically a sampler node to load some initial state) is identified with a double border, see figure~\ref{fig:SCCsystem}.}
\label{tab:legendN}
\end{figure}

\subsection{Container Nodes}

\textit{Container} nodes are partitioned into two subtypes: Particle container nodes and Environment container nodes. 

\paragraph*{Particle  container nodes:}mappings that take the node and the state of the system, and return the multiset of particles in that container at that state. When the system is in a particular state, the set of mappings of all the containers forms a partition of all the particles in the system. 

There are two types of particle container nodes: \textit{Samples} and \textit{Tanks}. Tanks are protected containers, in that particles inside them can not be edited. Particles in tanks can be moved in and out, but cannot be changed; any changes must be made over samples, so that the designer must explicitly decide what will be changing.

\noindent
{\it Examples:} A beaker being used for an experiment; a pipette; a petri dish.

\paragraph*{Environment container nodes:}similar to particle container nodes, except that they contain non-particle objects and information in the system. 
The system can have multiple environments, to make reference to the things in the environment easier. 
For example, one might want to store a time record separately from summary statistics or log information; one might want different local temperatures for different containers. These are all accessed via some mapping from the node and state of the system to the dynamic information and objects.
\noindent
{\it Examples:} Temperature readings; Bunsen burner; stirrer; observation log.\\

Container nodes are never directly connected to each other. 
All communication between them is mediated by control nodes. 
This means we always have control over the movement, similar to having valves and drip taps installed in normal chemistry glassware. 
We can allow things to flow through these controls freely, but we always have the option to restrict or stop any flows.

\subsection{Action Nodes}
\textit{Action nodes}, a kind of Control node, are where we actually modify particles through movement, linking, decomposition or any other change. Actions can modify particles only in a \textit{sample}. This means we always need to designate which particles we are changing before change occurs. This protects the particles in \textit{tanks}. 

\noindent
{\it Examples:} Concatenate strings; 
form chemical bond; execute an automata chemistry program string

\subsection{Admin Nodes: sampler, observer, decision}
\textit{Admin nodes}, kinds of Control node, are where particles and environments are moved and inspected.

\paragraph*{Sampler:}Information Admin nodes that move particles between particle containers (\textit{tanks} and \textit{samples}).

\noindent
{\it Examples:} Extracting a sample with a pipette for testing; choosing a neighbour to combine with the current particle.

\paragraph*{Observer:}Information Admin nodes that observe particles and/or environment state of other nodes.  They do not change any internal properties of particles or move them between containers. They can only see particles, derive information such as summary statistics, and modify the \textit{environment}.

\noindent
{\it Examples:} Taking notes in a log book; updating time in a discrete time system; updating the number of particles in the system.


\paragraph*{Decision:}Control Admin nodes used to change control flow. This is the only place control flow can branch, by using information the node has read from connected containers. 

\noindent
{\it Examples:} Triggering an event; continuing to the next phase; looping over a process; completing a time step; deciding to take a beaker off the heat.

\subsection{Termination Node}
\textit{Termination nodes}, a kind of Control node, are where execution of the AChem is explicitly terminated.
For an executable system, this implies the need for non-volatile memory for at least some containers, so that their contents can be inspected after a run; this is an implementation issue.

Not all graphs need to have an explicit termination node: we can define an \textit{open-ended} AChem that implicitly runs forever.


\subsection{Edges}
The nodes of our graphs are connected by edges capturing two kinds of relationship, Figure~\ref{tab:legendE}. 

\begin{figure}[t]

\begin{tabular}{cm{0.75\textwidth}}
\toprule
Element & Description\\
\midrule
Information Flow\\
\midrule
\begin{tikzpicture}
\node (A) at (0,0) [shape=rectangle,draw] {};
\draw[-, dashed] (A) -- (1,0);
\end{tikzpicture} & \textbf{Read}: allows reading of information from source container node (shown) in to target control node local state.\\[1mm]
\begin{tikzpicture}
\node (A) at (0,0) [shape=rectangle,draw] {};
\draw[-triangle 45, dashed] (A) -- (1,0);
\end{tikzpicture} & \textbf{Pull}: allows pulling of information out of  source container node (shown) by target control node.   Also allows reading from the source container.\\[1mm]
\begin{tikzpicture}
\node (A) at (0,0) [shape=rectangle,draw] {};
\draw[-triangle 45, dashed] (1,0) -- (A);
\end{tikzpicture} & \textbf{Push}: allows pushing (writing) of information from source control node local state into target container node (shown).  Also allows reading from the target container.\\
\midrule 
Control Flow\\
\midrule
\begin{tikzpicture}
\draw[-triangle 45] (0,0) -- (1,0);
\end{tikzpicture} & Solid arrow between 
control nodes indicates control flow in system.\\
\bottomrule
\end{tabular}
\caption{Graph edge types, and their graphical representation, used in MetaChem}
\label{tab:legendE}
\end{figure}

\paragraph*{Information Flow:}The first relationship marks movement of information between nodes. These relationships are always between container nodes and control nodes. Container nodes cannot directly transfer information; the same is true of control nodes. 

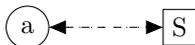
\begin{figure}[ht]
    \centering
    \begin{tikzpicture}
        \node (a) at (0,0) [shape=circle,draw] {a};
        \node (S) at (2,0) [shape=rectangle,draw] {S};
        \draw[-triangle 45, dashed] (a) -- (S);
        \draw[-triangle 45, loosely dotted] (S) -- (a);
    \end{tikzpicture}
    \caption{A combination of a read, a push, and a pull edge, abbreviated as a double-headed information edge}
    \label{fig:infoedge}
\end{figure}

\noindent
{\it Example.} Figure~\ref{fig:infoedge} show an action node that reads information from a sample, pulls a subset out of the sample (such as removing two particles to combine), performs its processing (such as combining the two particles), and then pushes (writes) the results back into the sample.

\paragraph*{Control Flow:}Our second relationship type marks the movement of control. These edges are between control nodes, and indicate what order we visit the control nodes. For most control nodes there can be only one outgoing control edge. The exception is decision nodes, whose purpose is to provide a branch point in control flow.

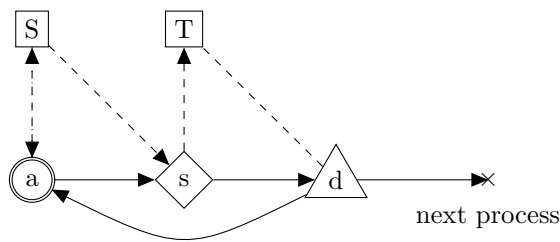
\begin{figure}[h]
    \centering
    \begin{tikzpicture}
        \node (a) at (0,0) [shape=circle, double, draw] {a};
        \node (S) at (0,2) [shape=rectangle,draw] {S};
        \draw[-triangle 45, dashed] (a) -- (S);
        \draw[-triangle 45, loosely dotted] (S) -- (a);
        \node (s) at (2,0) [shape=diamond, draw] {s};
        \node (T) at (2,2) [shape=rectangle,draw] {T};
        \node (d) at (4,0) [shape=regular polygon,regular polygon sides = 3,draw, label=center:d] {\phantom{.}};
        \node (end) at (6,0) [] {$\mathbf{\times}$};
        \node (text) at (6,-0.5) [] {next process};
        \draw[-triangle 45] (a) -- (s);
        \draw[-triangle 45] (s) -- (d);
        \draw[-triangle 45] (d) -- (6,0);
        \draw[-triangle 45] (d) .. controls (2,-1) .. (a);
        \draw[-triangle 45, dashed] (S) -- (s);
        \draw[-triangle 45, dashed] (s) -- (T);
        \draw[dashed] (d) -- (T);
    \end{tikzpicture}
    \caption{Example of a graph depicting actions and decisions.  Control starts in the action node \textbf{a} (the double-edged icon indicates it is the initial node in the system), which can pull and push particles from/to the sample node \textbf{S}.
    Next, control moves to the sampler node \textbf{s}, which can pull particles from the sampler node \textbf{S}, and push particles to the tank node \textbf{T}.
    Next, control moves to the decision node \textbf{d}, which has two control edges coming out of it.
    Based on the working of the decision node, using information it reads from the tank node \textbf{T}, control either loops back to the action node \textbf{a}, or continue on to the next process in the system.}
    \label{fig:controledge}
\end{figure}

\noindent
{\it Example.} Figure \ref{fig:controledge} shows the use of a decision node to control looping.
An action node links particles in the sample (as in the example shown in  figure~\ref{fig:infoedge}).
    Next he sampler node moves the content of the sample container to the tank. 
    Then the decision node checks if the system is finished with linking; if not, control loops back to the action node, to continue linking, otherwise control continues on to the next process in the system.

\subsection{Graph as an Executable Algorithm}
We have so far discussed separating out the parts of an AChem into nodes and   forming a graph using information and control flow edges between these nodes.

This graph is an executable script; it can be executed in software
\citep{Rainford2019-ALife}. During execution we have a single execution pointer\footnote{%
Later versions might support multiple threads of control.
} on a control node that executes a transition function, then follows the control edges, moving around the graph executing the transitions defined in the control nodes (see section~\ref{sec:static} for how the internal processes of nodes are defined).

In this version a graph is a static object that is defined before execution and remains unmodified by execution. This is the initial static graph form of MetaChem. 

Information flow edges can be seen as directing input and output of the nodes. In a way container nodes act like ``blackboard systems" \citep{Hayes-Roth1988-hp}, being constantly modified and updated by ``experts", the control nodes. Samples exist to section off part of a container and thereby to control which parts of our ``blackboard" each of our ``experts" can edit. In terms of a physical blackboard, they allow us to draw a box around the content of our tank and write ``Do not erase!" next to it.

\section{Descriptive levels}
\label{sec:overview}

The MetaChem graphical formalism allows a modular description of an AChem in term s of its subcomponents.
The level at which we define these subcomponents gives the \textit{descriptive level} of our graph.

\subsection{Expanding and summarising}

Moving between descriptive levels may expand nodes into subgraphs,
or summarise subgraphs as nodes. 

In the case of expansion, the resulting subgraph can still be described in terms of the component functions of the original single node. Such a graph can therefore be summarised in a well-defined manner as a single node. 

Starting with an arbitrary subgraph and summarising it into a single node is in general harder. 
If the subgraph we wish to describe as a single node can be broken up into the different component functions, then we can summarise it as a node with these functions. 
Failing this, we summarise as follows:
\begin{itemize}
    \item All information needed in the subgraph is read in during the node's read phase, so the new node has all the read connections to containers in the larger graph that exist in the subgraph.
    \item Any samples and variables are also taken; in some cases where there is sampling from a tank, this will need to occur in a separate sampler node. 
    This will give all the container connections needed. 
    \item We then perform all processing in the subgraph in a single action node, including observations.  
    \item Pushing samples to tanks requires another separate sampler node. 
\end{itemize}
So in the worst case we can summarise any subgraph as at most three nodes.

\subsection{Node names}

With these different levels and complex systems and multiple nodes of the same type, the basic single letter tags used before are not sufficient. In order to distinguish the different nodes, we use tags with two-part names. 

For containers we write \textbf{X:}label, where $\mathbf{X} \in \{\mathbf{T},\mathbf{S},\mathbf{V}\}$,
and for control nodes we write \textbf{x:}label, where $x \in \{\mathbf{s},\mathbf{o},\mathbf{d},\mathbf{a},\mathbf{t}\}$.
The type tag, \textbf{X} or \textbf{x}, is part of the overall name\footnote{%
A form of ``Hungarian'' notation.
}. 
The label is a non-empty string of alphanumeric characters and underscores.

We can use the same label for different types of containers, for example: \textbf{T:}particles and \textbf{S:}particles for a tank and a sample container of particles, or \textbf{S:}sample and \textbf{s:}sample for a sample container and its control node. 
In the first example we are labelling based on sort of content, which is the same for both the tank and the sample. 
In the second we label the function in the system, the container contains a sample and the sampler takes a sample. 
These labels can be used in the same graphs for different nodes, as the type is part of the node name.

Container nodes with the same name in a graph represent the same node: they may be drawn separately for clarity. 

Control nodes with the same name are not necessarily the same node, but do apply the same process to the data they read in: they have identical internal functions. However, they read in from different containers given by the information edges, and transition to a different node after they are completed, given by the control edge. Since they have no memory, control nodes with the same name and the same information and outgoing control edges are equivalent nodes, and could be replaced with a single node. 
Since the node receives no information directly from the previous control node, and can have multiple incoming control edges, these edges are not important for node equivalence.


\subsection{StringCatChem: an illustrative toy example}
To illustrate the use and power of MetaChem at different descriptive levels, we introduce StringCatChem, a ``toy" AChem. StringCatChem is simple and small enough to run by hand, and can be fully and succinctly decomposed into its parts. We use MetaChem to describe full scale AChems in later sections.

In StringCatChem the atoms are character strings, and composites are formed by string concatenation. StringCatChem is situated in a collection of well-mixed tanks. 
When a string is selected for reaction, it is checked if it contains any identical adjacent letters; if so it is split between them. If not, a second string is selected at random from the same tank, and the strings are concatenated. The split or combined strings remain in the same tank. In a separate operation, strings are also randomly transferred  between tanks.

StringCatChem is very simple,
just forming random stings with no double letters. 
It will continue to act until 
all the strings have matching letters at the starts and ends, or there is one large string. 
After this the system will not change, as any concatenation will be split apart again before another concatenation can occur. 
StringCatChem is therefore not a good choice of AChem if one wishes to study interesting behaviours such as replication, open-endedness or the transition to life. However, it makes a good illustration of MetaChem: the whole system can be implemented with four container nodes and 13 control nodes.

\subsection{Macro Level}
The \textit{Macro} level view provides an overview of the entire AChem.
It rarely deals with individual values, atoms or interactions in the AChem.  Even a generation or time step at this level is just an update process.

\subsubsection*{Macro-level StringCatChem}

\begin{figure}
\includegraphics[width=\textwidth]{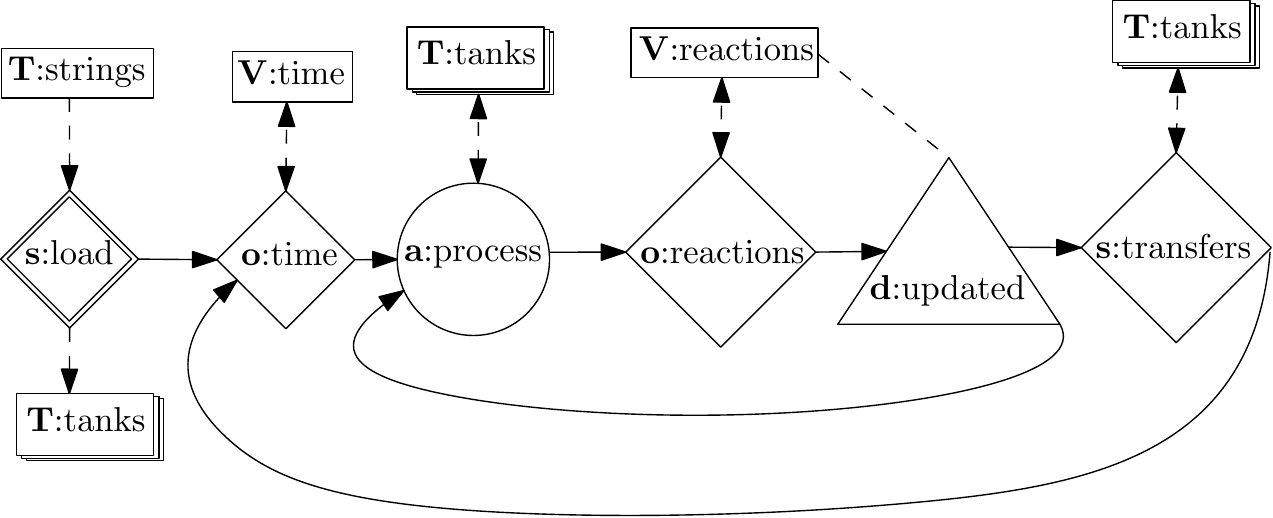}
\caption{Macro level description of (open-ended) StringCatChem.}
\label{fig:SCCsystem}
\vspace{3mm}
\includegraphics[width=\textwidth]{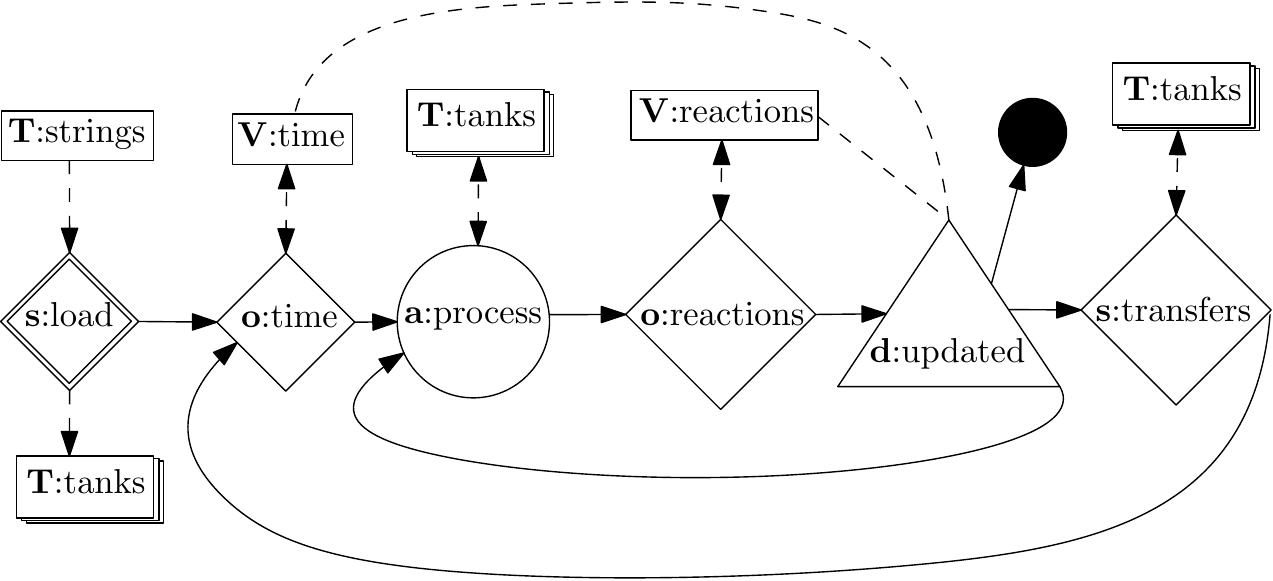}
\caption{Macro level description of explicitly-terminating StringCatChem.}
\label{fig:SCCsystemterm}
\end{figure}

Figure~\ref{fig:SCCsystem} shows the macro level description of StringCatChem. 
(A modification for explicit termination,
shown in Figure~\ref{fig:SCCsystemterm}, is discussed below.)
The start node is the control node with a double boarder. In many systems, for example that start with an initial set of particles to be loaded into the system as here, the start node will have no incoming control flow edges. 

\begin{figure}
\includegraphics[width=\textwidth]{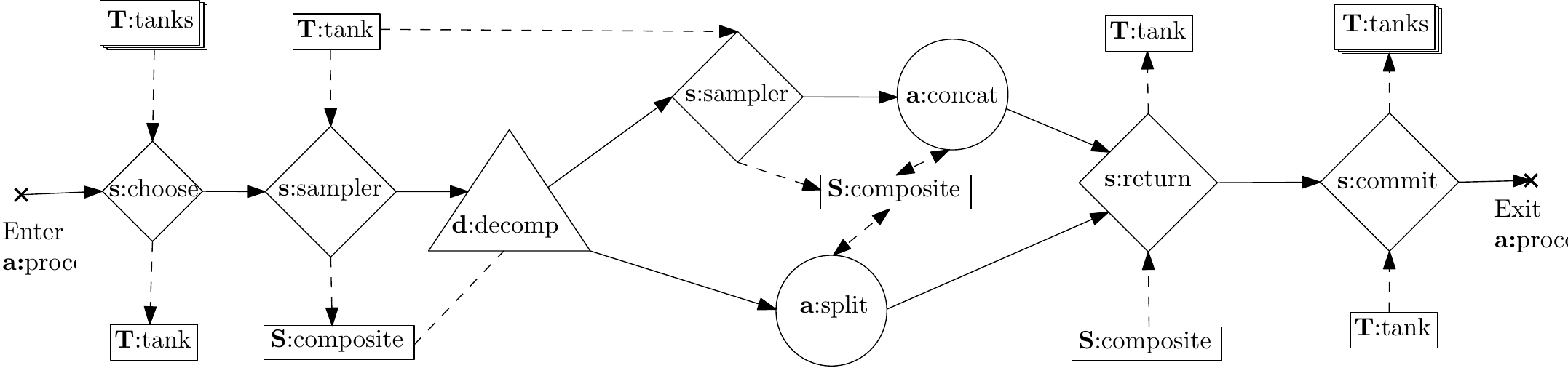}
\caption{Micro level description of the \textbf{a}:process node of the StringCatChem.}
\label{fig:SCCprocess}
\end{figure}

\begin{itemize}
    \item StringCatChem starts with the \textbf{s}:load node loading a set of strings into the set of tanks.
    \item The observer \textbf{o}:time then increments the time variable.
    \item The action \textbf{a}:process is responsible for the splitting and concatenation reactions that occur in the individual tanks. This is expanded later in Figure \ref{fig:SCCprocess}.
    \item The observer \textbf{o}:reactions then increments the reaction variable, to keep track of the number of reactions done in this timestep.
    \item The decision \textbf{d}:updated checks if the update cycle is complete (if enough reactions have been performed).  If not complete, it moves control back to \textbf{a}:process. If complete, it moves control on to \textbf{s}:transfers.
    \item The sampler \textbf{s}:transfers moves particles between tanks,
     then loops back to \textbf{o}:time for the next timestep.
\end{itemize}

This description gives us a high-level overview of the main operating loops of the system and of the set of significant processes. We can see  that there is a random unsynchronised update. Timing is discrete, and there are multiple tanks with movement between them. These are the kinds of elements and properties of a system that should be visible at the macro level.

This is the macro-level description of the system; it is an open ended system with no inherent termination point. For implementation purposes however we might add an explicit \textbf{V:}time dependant termination, Figure \ref{fig:SCCsystemterm}. Adding explicit termination to an open-ended system is usually done by adding a termination node to the decision at the end of the update loop, normally with the decision based on some variable, such as a time or generation variable, or anything else the designer wants to use to trigger termination of the run.

There is also a textual representation form for these graphs, which can be used for defining and executing them. An example of this textual form can be found in \cite{Rainford2019-ALife}.

\subsection{Micro Level}
The \textit{Micro} level view provides a focus on the actual action and effects on different particles and environments in the system. It can be thought of as the algorithm or pseudo-code level description of the AChem.

\subsubsection*{Micro-level StringCatChem of \textbf{a}:process }
As an example we expand the \textbf{a}:process node from figure~\ref{fig:SCCsystem} into a graph showing the internals of how this action occurs, Figure \ref{fig:SCCprocess}.

In summary, we choose a tank, then choose a particle string from it. We  decide (based on the presence of a double character) whether to decompose the string or not. In one case we split the string, in the other we sample a second string and concatenate them. The resulting particle string(s) are then returned to the same tank the original came from, and the tank is returned to the collection of tanks.
In terms of the graph, this is described as:
\begin{itemize}
\item sampler \textbf{s}:choose pulls the contents of a partition from the set \textbf{T}:tanks and pushes its contents into \textbf{T}:tank.
\item \textbf{s}:sampler pulls a particle from that \textbf{T}:tank and pushes it to \textbf{S}:composite
\item \textbf{d}:decomp decides if the particle can be decomposed or instead should be concatenated with another particle
\item if the decision is to decompose, control moves to \textbf{a}:split, which pulls the particle from \textbf{S}:composite, splits it, and pushes the resulting two particles back to \textbf{S}:composite
\item if the decision is not to decompose, control moves to \textbf{s}:sampler, which pulls another particle from tank \textbf{T}:tank and pushes it to \textbf{S}:composite; control moves to \textbf{a}:concat, which pulls the two particles from \textbf{S}:composite, concatenates them, and pushes the resulting particle back to \textbf{S}:composite
\item either path results in control being at \textbf{s}:return,
which pulls the resultant particle(s) from \textbf{S}:composite and pushes them into \textbf{T}:tank
\item finally, \textbf{s}:commit pulls the entire contents of \textbf{T}:tank and pushes it back into the same partition of \textbf{T}:Tanks which it originally come from.
\end{itemize}


\subsection{Physics Level}

The \textit{Physics} level view deals with the hard-coded details of implementation. It is the designer's choice what is the lowest level of detail needed; anything the designer considers to be unchangeable occurs at this level. 
This is the full program code level description, defining the internal processes of the control nodes.


\subsubsection*{Physics StringCatChem of \textbf{a}:split}

As an example we expand the \textbf{a}:split node from Figure \ref{fig:SCCprocess}, which splits a string that contains a double character.
In our implementation (available at \url{github.com/faulknerrainford/MetaChem.git}),
this is defined in the MetaChem package of Python,
by subclassing the ControlNode class to provide the specific implementation.
The ControlNode class defines the action of any control node in terms of components of the overall transition function,  Listing \ref{lst:transition}  (see Section~\ref{sec:static} for details).
\begin{Code}[tp]
\lstset{language=Python}
\lstset{frame=lines}
\lstset{caption=Transition function as defined in ControlNode class}
\lstset{label={lst:transition}}
\begin{lstlisting}
def transition(self):
    self.read()
    if self.check() < random.random():
        self.pull()
        self.process()
        self.push()
    pass
\end{lstlisting}
\end{Code}

The specific node subclass provides an implementation for each of these components in order to define the required processing,
 Listing \ref{lst:stringcatsplitaction}.
The specific implementation is defined as follows:

\begin{Code}[tp]
\lstset{language=Python}
\lstset{frame=lines, breaklines=true}
\lstset{caption=\textbf{a:}split node as described in Python using StringCatSplitAction. Action is a subclass of ControlNode class which contains the transition function. All aspects of Action are overwritten in StringCatSplitAction}
\lstset{label={lst:stringcatsplitaction}}
\begin{lstlisting}
class StringCatSplitAction(node.Action):

    def __init__(self, writesample, readsample, readcontainers=None):
        super(StringCatSplitAction, self).__init__(writesample, readsample, |\Suppressnumber|
            readcontainers) |\Reactivatenumber|
        self.sample = None
        pass

    def read(self):
        self.sample = self.readsample.read()

    def check(self):
        return super(StringCatSplitAction, self).check()

    def pull(self):
        self.readsample.remove(self.sample)

    def process(self):
        doubleindex = [i for i in range(0, len(self.sample) - 1) if |\Suppressnumber|
            self.sample[i] == self.sample[i+1]] |\Reactivatenumber|
        index = random.choice(doubleindex)
        self.sample = [self.sample[0:index], self.sample[index:0]]
        pass

    def push(self):
        self.writesample.add(self.sample)
\end{lstlisting}
\end{Code}


\begin{description}
\item[read()] Read the contents of the attached container(s) ({\tt readsample})  into local state. Here {\tt read\-sample} is \textbf{S:}composite, defined by the graph topology on system setup.  The local state is {\tt self.sample}. 
\item[check()] Check that we want to continue processing.  Here we use a default check action that returns {\tt 0}, so the action will always occur. All checks needed have already been made in the previous decision node \textbf{d}:decomp (figure~\ref{fig:SCCprocess}), so the action here is deterministic.
\item[pull()] Pull the relevant particles out of the attached container(s). Here the sample should contain only one particle, so we then remove that particle from \textbf{S:}composite, using the container's {\tt remove()} function. 
We specify which particle to remove in terms of the local state of the control node; hence that particle must have been read from the attached container earlier.
\item[process()] Process the material in the local state. Here, process function  finds the double letter, and splits the string at that point into two particles/strings, overwrite the internal sample state.
\item[push()] Push the relevant particles into the attached container(s).
Here we push the two split strings into the {\tt writesample} container, using the container's {\tt add()} function.
Here {\tt writesample} is \textbf{S:}composite, defined by the graph topology on system setup.
The graph topology could later be modified so that {\tt writesample} referred to a different container, without having to change the implementation here.
\end{description}

In the case of most processes the lower level instructions will use at least some default functionality; in this case it is the check function. In the case of other sorts of control nodes such as samples it might be the process function. These defaults are defined in the relevant superclass code. 

All interactions with containers are mediated through the interface of the containers' built in read(), add() and remove() functions. This allows the control node design to remain independent of the exact implementation of the containers. With this an AChem designer can build and test a system on the small scale using easy-to-manage list containers, and when they wish to scale up, they can reimplement the containers to use a database, without having to change the graph or the control nodes' implementations.

\subsection{Abstraction levels}
The abstraction levels are not restricted to these three levels: there can be systems made up of systems \citep{Rainford2018-xx} defined using additional levels.
It is up to the designer or modifier of the AChem to choose the abstraction level for  what is needed and useful in order to properly express and illuminate a particular system.

\section{Static Graph MetaChem}\label{sec:static}

Here we describe the internal structures of the nodes, in terms of the actions they perform.
We provide a mathematical specification in appendix~\ref{sec:maths}. 
This is Static Graph MetaChem: none of the actions described here change the structure of the graph.
Future versions will include actions that can dynamically change the structure of the graph as the algorithm executes.


\subsection{Control nodes and edges}
The control flow defines the AChem's algorithm: 
evaluate the current control node's definition, then move to the next control node, and repeat.
In the implementation, it executes the current node's transition function, which (potentially) changes state and then moves on to the next node; by traversing the graph in this manner it performs the relevant computation. The is no automatic termination rule on these systems as chemistries don't technically ever stop but for a particular algorithm we can define a number of transitions we will perform before stopping. We could also provide a control node with no outgoing edge. This would force termination.


All control nodes have the same basic structure for their state transition function, 
defined through component transition functions: \textit{read}(), \textit{check}(), \textit{pull}(), \textit{process}(), \textit{push}(), \textit{next}(), executed sequentially:
\[
transition = read \fatsemi check \fatsemi pull \fatsemi process \fatsemi push \fatsemi next
\]
where $\fatsemi$ indicates sequential ordering of function application from left to right.

Each of these component functions plays a different role in the transition and thus uses a different aspect of the state.
\begin{description}
\item[\textit{read}] 
Collect information from (connected) external containers into temporary local containers, for used by the following functions. This action does not modify the external containers; it copies the relevant particles and values into temporary local state.

\item[\textit{check}] 
Generate a threshold probability value $p$ from local state information.
This $p$ is used to determine if the rest of the component functions (the ones that actually alter containers' contents) occur. 
In the current implementation, it generates $p$ from its local state, then generates a uniform random number $r$; if $p< r$, execution continues, otherwise it moves directly to the $next$ node. The shape of $p$ can be defined such that the probability of the process follows the desired distribution.

\item[\textit{pull}] 
Remove particles and change information in external containers. Any information so removed must already have been copied to local containers by the earlier \textit{read}(), where it is available for local processing; such a $read$ followed by $pull$ has the effect of moving the particles or information. However, read information does not have to be pulled: it can be copied, rather than moved.

\item[\textit{process}] 
Perform the main computation for the node.
This is where the ``chemistry'' happens. It modifies the state of local particles and variables, including creating new particles and variables and destroying old ones. 

\item[\textit{push}] 
Copy particles and values from local state into external containers. 

\item[\textit{next}] 
Wipe the local state and move control to the next node. 
All control nodes except decision nodes have exactly one outgoing control edge, 
so the move is deterministic.
For decision nodes, $process$ designates the target node, and stores it in local state. This is used by $next$ to move to the chosen next node. 
\end{description}

These component functions operate on the node's local state, which exists only for the duration of the overall transition. 
Local particle containers and local environment containers are destroyed once the transition function is completed, so control nodes have no lasting state or memory. Any information used by a control node must come from containers at the start of a transition by using \textit{read}(); any local information or objects that need to remain in the system must be written back to containers by \textit{push}().

\begin{figure}[tp]
\centering
\includegraphics[trim = 0 0 17mm 0, clip, width=0.58\linewidth]{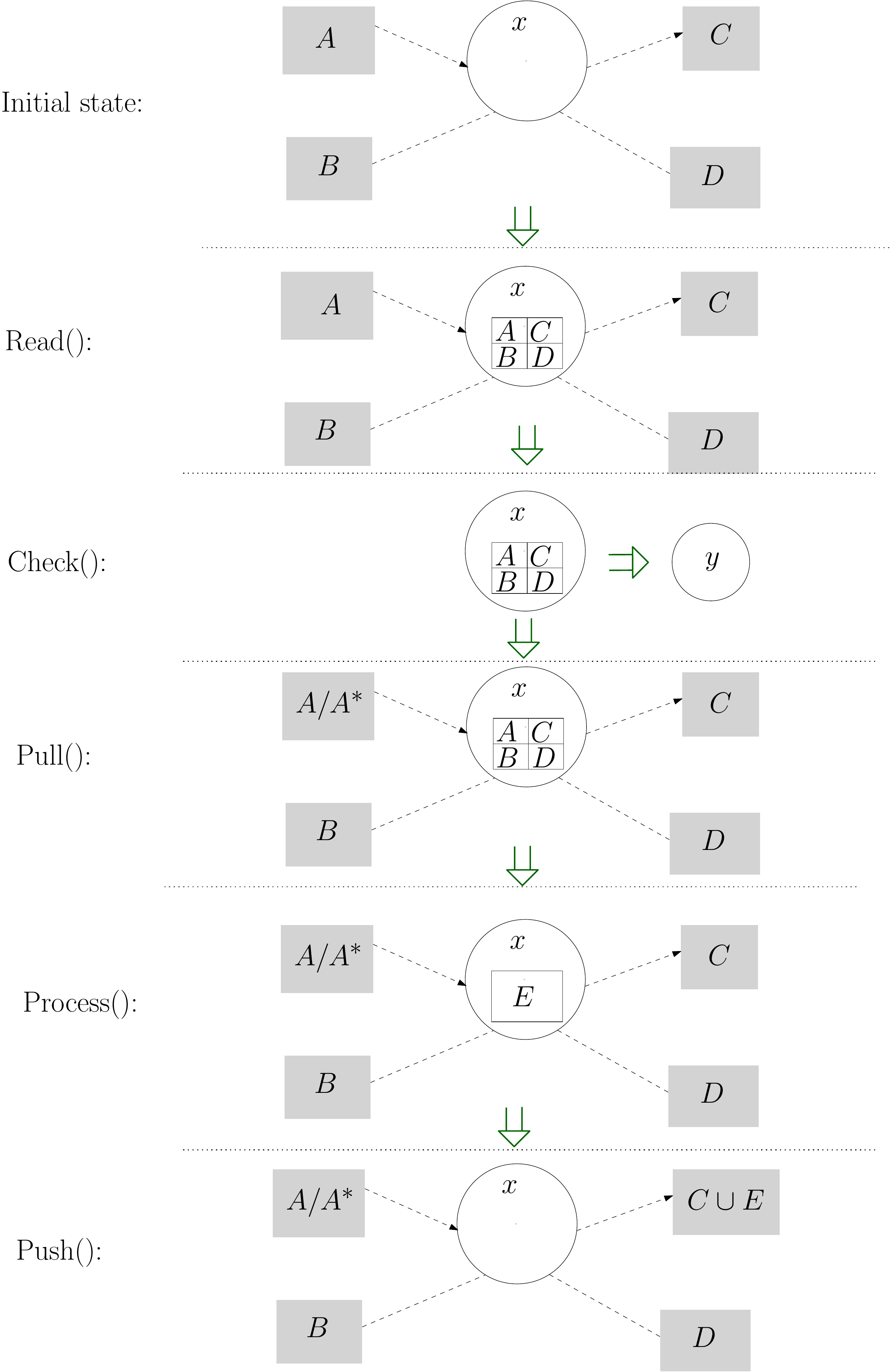}
\caption{Summary of movement and processing of information done by transition functions in a node}
\label{fig:transsum}
\end{figure}

These functions are summarised in Figure~\ref{fig:transsum} and discussed in the context of specific node types below.

\subsection{Control node subtypes}

Control nodes are partitioned into subtypes: action, decision, sample, observer, termination. 
We distinguish these node subtypes by requiring some of their transition function components to have no effect (to be the null operation, or the identity transformation), or by limiting the types of containers they can interact with during the transition, Table~\ref{tab:transitionfunction}.
The constraints on these subnodes help control the complexity of the system definition.

\begin{table}[tp]
    \centering
    \begin{tabular}{lccccc}\toprule
        & action & decision & sample & observer  & termination\\
        \midrule
        read & \checkmark & \checkmark & \checkmark & \checkmark \\
        check & \checkmark & & & \\
        pull & \checkmark & & \checkmark & \checkmark \\
        process & \checkmark & \checkmark & & \checkmark \\
        push & \checkmark & & \checkmark & \checkmark \\
        \bottomrule
    \end{tabular}
    \caption[Transition functions used by control nodes]{Transition functions used by different types of control nodes; unchecked functions always use their default behaviour}
    \label{tab:transitionfunction}
\end{table}

\begin{table}[tp]
    \centering
    \begin{tabular}{lccccc}\toprule
        & action & decision & sample & observer  & termination \\
        \midrule
        tank & & & \checkmark & \\
        sample & \checkmark & & \checkmark & \\
        environment & \checkmark & & & \checkmark \\
        \bottomrule
    \end{tabular}
    \caption[Accessible container nodes for each control node]{The  container nodes that can be modified by a control node (by \textit{push} or \textit{pull}); \textit{read} can be performed on any container}
    \label{tab:infolimits}
\end{table}

\paragraph{Action :} read in information, check if an interaction occurs, process the particles in the system for the reaction to happen.
This node type is not limited in which transition functions it executes, but it is limited in which containers it can \textit{push}() to, Table \ref{tab:infolimits}. 
The limit to modify only samples allows parallelisation, and encourages controlled modification. The designer is required to consider what they wish to modify before they modify it, as they must first \textit{sample} it from the tanks.

\paragraph{Decision :} process the information from its containers and return a choice of the possible next control nodes. It is limited to \textit{read}() and \textit{process}(), so it cannot change the contents of any of the containers.  

\paragraph{Sampler :} move particles between containers. It does not compute or process information, and it does not modify any particles or environment variables.
It is therefore limited to \textit{read}(), \textit{pull}() and \textit{push}(). 

\paragraph{Observer :}  observe but do not modify particles; modify the environment. 
It can \textit{read}() to view containers; it can \textit{pull}() to edit only environment variables. It can \textit{process}(), to compute summary statistics and changes to the environmental variables, and it can \textit{push}() to commit those changes back to the environment.

\paragraph{Termination :}  terminate execution.  It does nothing, so does not use any of the component transition functions.

\subsection{Container nodes}

Container nodes are interfaces between control nodes and the things in the system, rather than control elements themselves. We prevent any modification of objects inside container nodes, to preserve this separation. 
Every container node has three functions forming its uniform interface: \textit{read}(), \textit{add}() and \textit{remove}(). 
These are used respectively by the \textit{read}(), \textit{push}() and \textit{pull}() functions of control nodes. 
Internal data can be organised in any way the node designed sees fit as long as it provides these three functions. 
An implementation could move from using a list to a database by changing only the container, and not need to make any change to control nodes using it.

Container nodes are partitioned into two subtypes: Particle container nodes and Environment container nodes. 

\paragraph{Particle container node:} 
contains a multiset (bag) of particles; the contents of this multiset changes as the AChem executes. 
The state of all the containers in the system partitions  the particles in the system. 
There are two sub types of particle container nodes: Samples and Tanks. Tanks are protected containers. Particles in tanks can be moved in and out but cannot be changed in the tank; any changes must be made in Sample containers, so the designer has to decide what will be changing.

\noindent
{\it Examples :} A beaker being used for an experiment, a pipette, a  petri dish.

\paragraph{Environment container node:}  contains non-particle objects and information in the system. 
The system can have multiple environments, to make reference to the things in the environment easier. For example, one might want to store a time record separately to summary statistics or log information. These are all still accessed via a mapping from the node and state of the system to the dynamic information and objects.

\noindent
{\it Examples :} Temperature readings, Bunsen burner, stirrer, observation log.\\



\noindent
The limits on access to containers placed on control nodes is given in Table~\ref{tab:infolimits}. 
Any control node can \textit{read}() any container node (information is always knowable). However, we limit the modification of containers to certain types of control nodes to make it easier to track activity in the system. This should also encourage limiting the scope of individual nodes to a basic action that may be reusable.

\subsection{Examples from StringCatChem}
\subsubsection{Local state}
Here we give examples of the behaviour of some of the control nodes in StringCatChem. 
These involve reading particles and environment variables into local state.
This local state is modelled (see appendix, eqn.\ref{eqn:localstate})  as a pair of mappings, the first from particle (tank and sampler) node names to contents, the second from environment node names to contents.

\subsubsection{Sampler node}
Here we describe the micro level \textbf{s}:sampler from Figure~\ref{fig:SCCprocess}.
Sampler nodes have \textit{read}(), \textit{pull}() and \textit{push}() functions, Table \ref{tab:transitionfunction}.
In this example, this sampler randomly selects a single particle from a tank to move to a sample container. 

\paragraph*{read():}Read the contents of the containers attached by information edges.
The node \textbf{s}:sampler has two read edges, one to \textbf{T:}tank (also a pull edge) and one to \textbf{S:}composite (also a push edge).
After the read, the local state $\Lambda$ has a copy of the states of these two particle containers (defined in the global state $G$); there are no connected environment containers, so it has an  empty environment component:
\begin{align*}
 &G = ( \{ \mbox{\textbf{S:}composite}\mapsto\Sigma , 
    \mbox{\textbf{T:}tank}\mapsto T, \ldots \}, \{  \ldots \} ) \\
 &\Lambda = ( \{ \mbox{\textbf{S:}composite}\mapsto\Sigma , 
    \mbox{\textbf{T:}tank}\mapsto T\}, \{ \} )
\end{align*}
where $\Sigma$ is (a copy of) the particles in \textbf{S:}composite,
and $T$ is (a copy of) the particles in \textbf{T:}tank.

\paragraph*{pull():} Select a random particle $\tau$ from $T$, and delete (pull) the corresponding particle from the external container \textbf{T:}tank\footnote{%
For notational simplicity in these examples, we assume here that containers contain \textit{sets} of particles; in the full formalism, the containers contain \textit{multi-sets} (bags) of particles (appendix~\ref{app:maths:state}), allowing multiple instances of a given species.
}.
\begin{align*}
 &G = ( \{ \mbox{\textbf{S:}composite}\mapsto\Sigma , 
    \mbox{\textbf{T:}tank}\mapsto T\setminus \{ \tau \}, \ldots \}, \{  \ldots \} ) \\
 &\Lambda = ( \{ \mbox{\textbf{S:}composite}\mapsto\Sigma , 
    \mbox{\textbf{T:}tank}\mapsto T\}, \{ \} )
\end{align*}

\paragraph*{push():} Push the selected particle $\tau$ to the \textbf{S:}composite sample.
\begin{align*}
 &G = ( \{ \mbox{\textbf{S:}composite}\mapsto\Sigma \cup \{ \tau \} , 
    \mbox{\textbf{T:}tank}\mapsto T\setminus \{ \tau \}, \ldots \}, \{  \ldots \} ) \\
 &\Lambda = ( \{ \mbox{\textbf{S:}composite}\mapsto\Sigma, 
    \mbox{\textbf{T:}tank}\mapsto T\}, \{ \} )
\end{align*}
On moving to the next control node, the local state $\Lambda$ is destroyed, 
with the overall result that particle $\tau$ has moved from the sampler to the tank.

\subsubsection{Observer node}
Here we describe the macro level \textbf{o}:time from Figure~\ref{fig:SCCsystem}.
Observer nodes have  \textit{read}(), \textit{pull}(), \textit{process}() and \textit{push}() functions, Table \ref{tab:transitionfunction}. 
In this example, this basic observer increments a variable representing time.
\paragraph*{read():}Read the contents of the containers attached by information edges.
The node \textbf{o}:time has one read edge (also a pull and a push edge), to the environment container \textbf{V}:time. 
After the read, the local state $\Lambda$ has a copy of the state of this environment container; there are no connected particle containers, so it has an  empty particle component:
\begin{align*}
 &G =  ( \{ \ldots \} , \{ \mbox{\textbf{V}:time}\mapsto V, \ldots \}    )
 \\
 & \Lambda = ( \{ \} , \{ \mbox{\textbf{V}:time}\mapsto V \}    )
\end{align*}
where $V$ is (a copy of) the environment in \textbf{V}:time.
Here, the environment contains a single variable, representing the time.

\paragraph*{pull():}As our \textbf{V:}time container only contains a single variable our pull function clears the \textbf{V:}time container. 
\begin{align*}
 &G =  ( \{ \ldots \} , \{ \mbox{\textbf{V}:time}\mapsto \varnothing, \ldots \}    )
 \\
 & \Lambda = ( \{ \} , \{ \mbox{\textbf{V}:time}\mapsto V \}    )
\end{align*}

We remove the value from the attached container, because interactions with the container are limited to read(), add() and remove() functions; if we wish to update or modify a variable we must read it in, remove it from the container, then add the new version. 
If we simply push/add the new version without clearing the old one, behaviour is undefined and will depend on implementation.

\paragraph*{process():}This observer is a \textit{counter observer}: it increments a single variable. In this case the variable is time and the increment is 1. This is performed on the variable $V$ in local storage.
\begin{align*}
 &G =  ( \{ \ldots \} , \{ \mbox{\textbf{V}:time}\mapsto \varnothing, \ldots \}    )
 \\
 & \Lambda = ( \{ \} , \{ \mbox{\textbf{V}:time}\mapsto V+1 \}    )
\end{align*}

\paragraph*{push():}Push the incremented local variable back out into the \textbf{V:}time container for storage.
\begin{align*}
 &G =  ( \{ \ldots \} , \{ \mbox{\textbf{V}:time}\mapsto V+1, \ldots \}    )
 \\
 & \Lambda = ( \{ \} , \{ \mbox{\textbf{V}:time}\mapsto V+1 \}    )
\end{align*}
On moving to the next control node, the local state $\Lambda$ is destroyed, 
with the overall result that the environment variable \textbf{V}:time has been incremented by one.

\subsubsection{Action node}
Here we describe the micro level \textbf{a}:split from Figure~\ref{fig:SCCprocess}.
Action nodes can use all five component transition function. In the case of \textbf{a:}split we explicitly use some of these functions, and use the default definition of the others.

\paragraph*{read():} The \textbf{a:}split action node has an information edge to only one other node, in the form of a read, pull and push edge between it and \textbf{S:}composite. 
At this stage \textbf{S:}composite holds a single particle string, which  contains a double letter.
This string is read into the local particle container.
\begin{align*}
 &G = ( \{ \mbox{\textbf{S:}composite}\mapsto \{ \mbox{``prexxpost''} \} , \ldots \}, \{  \ldots \} ) \\
 &\Lambda = ( \{ \mbox{\textbf{S:}composite}\mapsto \{ \mbox{``prexxpost''} \} \}, \{ \} )
\end{align*}

\paragraph*{check():}In this particular system reactions are deterministic, so we use the default behaviour of the check function, which is to return {\tt 0}.
So the check test is true, and the rest of the action happens.

\paragraph*{pull():} Delete (pull) the string from the \textbf{S}:composite node.
\begin{align*}
 &G = ( \{ \mbox{\textbf{S:}composite}\mapsto \varnothing , \ldots \}, \{  \ldots \} ) \\
 &\Lambda = ( \{ \mbox{\textbf{S:}composite}\mapsto \{ \mbox{``prexxpost''} \} \}, \{ \} )
\end{align*}

\paragraph*{process():} Process the contents of the local state.
Here, divide the string at the double letter, 
and store the two resulting strings in the local particle state.
\begin{align*}
 &G = ( \{ \mbox{\textbf{S:}composite}\mapsto \varnothing , \ldots \}, \{  \ldots \} ) \\
 &\Lambda = ( \{ \mbox{\textbf{S:}composite}\mapsto \{ \mbox{``prex''}, \mbox{``xpost''} \} \}, \{ \} )
\end{align*}

\paragraph*{push():}Push the resulting strings back into the \textbf{S}:composite sample node. 
\begin{align*}
 &G = ( \{ \mbox{\textbf{S:}composite}\mapsto \{ \mbox{``prex''}, \mbox{``xpost''} \} , \ldots \}, \{  \ldots \} ) \\
 &\Lambda = ( \{ \mbox{\textbf{S:}composite}\mapsto \{ \mbox{``prex''}, \mbox{``xpost''} \} \}, \{ \} )
\end{align*}
On moving to the next control node, the local state $\Lambda$ is destroyed, 
with the overall result that \textbf{S}:composite sample node now contains the split strings.


\subsubsection{Decision node}
Here we describe the micro level \textbf{d:}decomp from Figure~\ref{fig:SCCprocess}.
Decision nodes use two of the five transition functions. They do not change the state of any of the containers (so no pull or push).
They just \textbf{read} containers, and compute (\textit{process})  the next control node to move to. 
\paragraph*{read():}The \textbf{d:}decomp decision node has a read edge to  between it and \textbf{S:}composite. 
At this stage \textbf{S:}composite holds a single particle string, which  may or may not contain a double letter.
This string is read into the local particle container.
\begin{align*}
 &G = ( \{ \mbox{\textbf{S:}composite}\mapsto \{ ``p_1p_2..p_n\mbox{''} \} , \ldots \}, \{  \ldots \} ) \\
 &\Lambda = ( \{ \mbox{\textbf{S:}composite}\mapsto \{ ``p_1p_2..p_n\mbox{''} \} \}, \{ \} )
\end{align*}

\paragraph*{process():}This function performs the computation that makes the decision. It returns one of the possible next control nodes, 
\textbf{s:}sampler or \textbf{a:}split\},
and stores this in the local state for the \textit{next} function to access.
\begin{align*}
 &G = ( \{ \mbox{\textbf{S:}composite}\mapsto \{ ``p_1p_2..p_n\mbox{''} \} , \ldots \}, \{  \ldots \} ) \\
 &\Lambda = ( \{ \mbox{\textbf{S:}composite}\mapsto \{ ``p_1p_2..p_n\mbox{''} \} \}, \{ \mbox{\textbf{V:}\_local} \mapsto c\} )
\end{align*}
where
\[c=
\begin{cases}
\mbox{\textbf{a:}split} & \text{if } \exists\, i \in 1..n-1\,|\,p_i = p_{i+1} \\
\mbox{\textbf{s:}sampler} & \mbox{otherwise}  
\end{cases}
\]
On moving to the next control node, $c$, the local state $\Lambda$ is destroyed, 
with the overall result that no containers have changed state, and the control node is the relevant one for the string in container \textbf{S:}composite.


\section{Jordan Algebra Artificial Chemistry}
\label{sec:jaachem}

MetaChem can be used to design new AChems, and to describe existing AChems.
Here we use MetaChem to describe 
our earlier Jordan Algebra Artificial Chemistry, JA-AChem \citep{Faulkner2016-sz}. We choose this as one extreme of an artificial chemistry. It is subsymbolic in that its bonding properties arise from the internal structure of its particles, and is designed to work at the level of atoms. It has both a linking action and a destructive action, which makes it more complicated in terms of algorithm than the StringCatChem already described.


\subsection{Overview of particles and linking}

Hermitian matrices provide a rich variety of properties such that we can use them as prime material for creating a subsymbolic AChem \citep{Faulkner2018-ex}, where emergent properties of the matrices dictate the linking capabilities/probabilities of a particle, and the algebra gives the structure of the composite particles. 
Here we give a condensed overview of the properties used to form atoms and composite particles;
in the next section we use these definitions in the overall MetaChem description of the JA-AChem.

\subsubsection{Definitions and properties}
The complex conjugate of the transpose of a matrix is written as $M^\dagger$.
A matrix $M$ is Hermitian if it is equal to the complex conjugate of its transpose:
$M = {M}^\dagger$.

The Jordan product of two square matrices is 
\begin{equation}\label{eqn:jordanprod}\textstyle
A \circ B := \frac{1}{2}(AB +BA)    
\end{equation}
Hermitian matrices are closed under the Jordan product \citep{McCrimmon2006-iu}.

The eigenvalues $\lambda$ and eigenvectors $\mathbf{v}$ of a square matrix $M$ are solutions to $(M-\lambda I)\mathbf{v} = 0$.
An $n$D matrix has $n$ (possibly degenerate) eigenvalues, and $n$ corresponding eigenvectors.
The eigenvalues of a Hermitian matrix are all real.

\subsubsection{Atoms}
The atoms in the Jordan Algebra AChem used here are specific $3\times 3$ Hermitian matrices. The atom set is:
\begin{equation}
\label{equ:atoms}
\left\{ A=
\begin{pmatrix}
x_{11} & x_{12} & x_{13}\\
x_{21} & x_{22} & x_{23}\\
x_{31} & x_{32} & x_{33}
\end{pmatrix}
: x_{ij} \in \{\pm 1, \pm i, \pm 1 \pm i, 0\}\right\}
\end{equation}
We use the eigenvalues of the matrices to define linking properties.

There are 14574 atoms, with 66 different sets of eigenvalues. 
We have many options and many different sorts of operations and linking behaviours are possible. 

\subsubsection{Composite particles}
We use unit eigenvectors $\hat{\mathbf{v}}_i$ and their corresponding normalised eigenvalues $\mu_i$ to define linking probabilities:
\begin{equation}\label{eqn:normeval}
	\mu_i = \lambda_i / {\textstyle \sum} \lambda_j
\end{equation}
We normalise the eigenvalues to ensure sensible linking probabilities of larger composites. (The sum of eigenvalues equals the trace of the matrix, which is required to be non-zero in this system.)

We defined the \textit{alignment} of two eigenvectors (one from each particle's matrix) as:
\begin{equation}\label{eqn:align}\textstyle
a_{i j}=\left(1-\frac{1}{2}((\hat{\mathbf{v}}_{i} . \hat{\mathbf{v}}_{j})+1)\right)
\end{equation}
Alignment uses the dot product between unit vectors,
which is the cosine of the angle between them.
Hence the alignment has a value between 0 and 1, being 0 if the vectors are parallel and 1 if they are anti-parallel. 

When linking two particles $A$ and $B$,
we calculate the normalised eigenvalues $\mu_{A_i}, \mu_{B_j}$ and the corresponding unit eigenvectors $\hat{\mathbf{v}}_{A_i}, \hat{\mathbf{v}}_{B_j}$.
We calculate all the alignments between pairs of eigenvectors, one from each particle.
We choose the highest value alignment (most anti-parallel eigenvectors) as the linking eigenvectors $(i,j)$.

We calculate the \textit{strength} of this alignment, using the corresponding eigenvalues: 
\begin{equation}\label{eqn:linkstrength}
s_{A_i B_{j}} =\mathcal{N}( \mu_{A_i}-\mu_{B_j}) 
\end{equation}
where $\mathcal{N}(x)$ is the value of the probability density of the normal distribution ($\mu=0$, $\sigma=1$) at $x$. 
This will give a probability of linking that is larger for more similar normalised eigenvalues. The normal distribution is not the only option we could use here;
we investigate other options in \cite{Faulkner2017-ap}.

We calculate the probability of linking based on the strength of the link and its alignment.  
\begin{equation}
\label{eqn:linkprob}
    p_{AB} = s_{A_i B_{j}}a_{A_i B_{j}}
\end{equation}

If the link is formed, the resulting composite particle is the Jordan product of the components.

\subsection{Macro description of Jordan Algebra Artificial Chemistry}
In addition to these particles,
we need an algorithm for how our system works. 
This covers not only the linking and decomposition aspects, but the entire behaviour of our chemistry. We define this using MetaChem, starting at the Macro level, which describes the overall behaviour of the system over time. We then look in more detail at the linking process. For more information on how decomposition works see \cite{Faulkner2016-sz}.

\begin{figure}
\includegraphics[width=\textwidth]{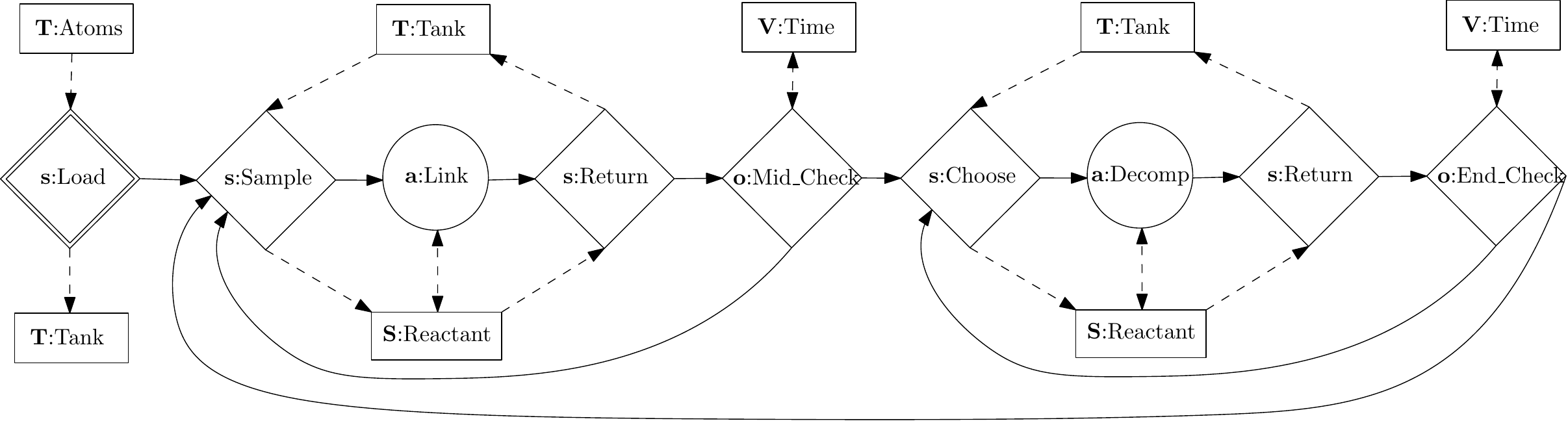}
\caption{Macro level description of JA-AChem operating over link and decomp loops}
\label{fig:JAMacro}
\end{figure}


The algorithm loads the initial atom or particle set and then operates over two loops (figure~\ref{fig:JAMacro}). These two loops are similar, starting with sampling from the tank followed by an operation. The first loop performs linking; the second loop performs decomposition.  The loops finish by returning their modified samples to the tank, updating timing variables, and checking if enough operations or time has passed to say whether the loop continues or if the system moves to the other loop.

If we make observations of our system, we add them to our logger, which can be added to the system in figure~\ref{fig:JAMacro}. The logger pushes to an external environment which is never pulled from. These observations can provide many different summary statistics. In later examples relating to linking and probabilities we observe and log: number of atoms in each particle, number of different atoms in each particle, size of particle trace, weight of particles, size of largest link in particles. 

This macro system level description does not cover the internal workings of our link and decomposition nodes. The algorithms for these are described in detail in \cite{Faulkner2016-sz}.

\subsection{Micro level description of linking in JA-AChem}

\begin{sidewaysfigure}
    \centering
    \includegraphics[width=\textwidth]{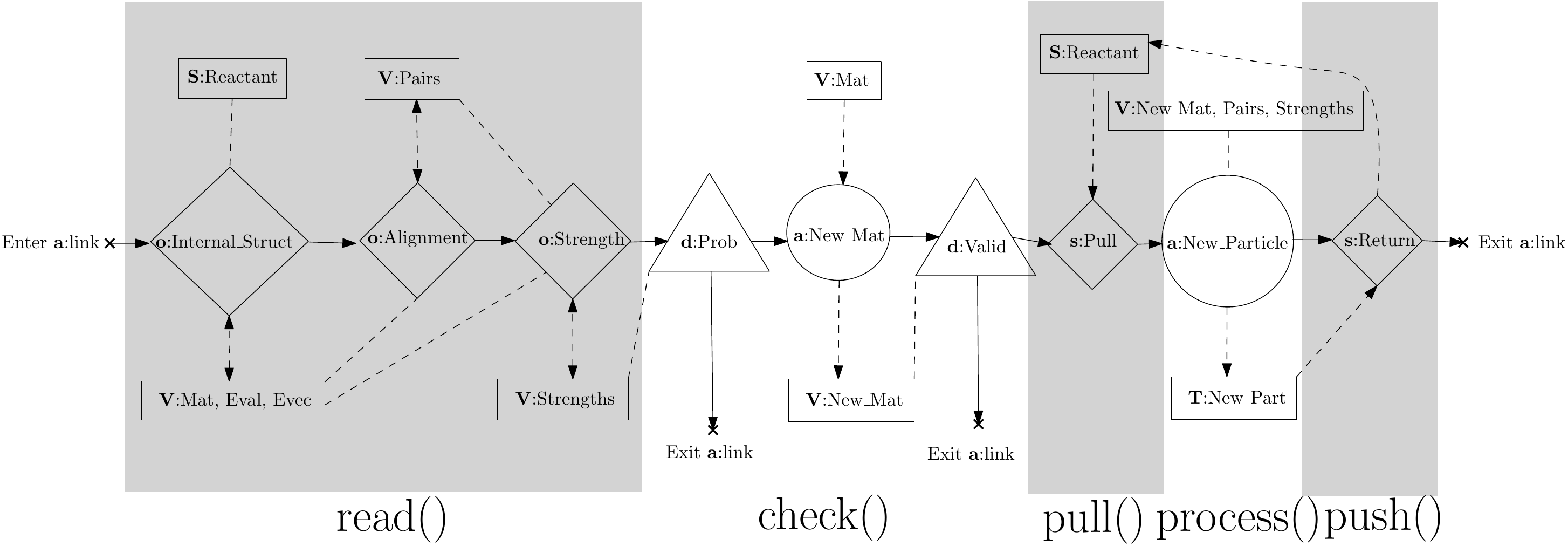}
    \caption[Micro Level description of \textbf{a:}link node from Figure \ref{fig:JAMacro}]{Micro Level description of \textbf{a:}link node from Figure \ref{fig:JAMacro}. Below are the labels for the grouping of control nodes according to transition function in the the outer node. The only external contain used here is \textbf{S:}Reactant. All other containers are part of $L_V$ and are lost at completion of control flow. We have no need here for $L_B$}
    \label{fig:JAACalink}
\end{sidewaysfigure}

Now we have the wider view of how this AChem works, we look in more detail at the micro level description of \textbf{a:}Link. This is defined by the MetaChem graph in Figure \ref{fig:JAACalink}. It has been defined and labelled in terms of the macro-level component transition functions in \textbf{a:}Link. We discuss the process in terms of these components below.

The \textbf{a:}Link action node uses all five component transition functions. The read, pull and push functions are all performed from and to the 
\textbf{S:}reactant sample. More interesting are the check and processing functions in this case. To describe these actions in more detail, we expand the \textbf{a:}Link action node into a micro level subgraph of the macro system. This means that all five transition functions are themselves defined in terms of micro-level graphs, and we introduce various explicit micro-level containers \textbf{V}:xxx, \textbf{T}:xxx, \textbf{S}:xxx to implement the macro-level local state.

Some of these subgraphs do processing where the high level component function does not. For example, processing does not occur in the macro-level \textit{read}() function, but does in the corresponding part of the micro-level graph. Now that we are taking a lower level view of this function, we can reveal more of the implementation, and with it the shortcuts we take to reduce repetition of calculations and have a smoother flow. So this lower-level description is not a formal refinement of the individual component functions, but involves some refactoring (rearranging) of the functionality.

\subsubsection{Expansion of macro-level \textit{read}()}
Three observers, \textbf{o:}internal\_struct, \textbf{o:}Alignment and \textbf{o:}Strength, gather the information needed for the linking probability check. 
They do this by reading information from \textbf{S:}reactants and processing it. 
The information about the internal structure and other derived values is stored in various environment containers \textbf{V}:Mat, \textbf{V}:Eval, \textbf{V}:Evec, \textbf{V}:Pairs, \textbf{V}:Strengths. 

The observer \textbf{o:}internal\_struct  pulls any previous matrices, vectors and eigenvalues from the container to clear it. It then extracts the matrix, normalised eigenvalues and unit eigenvectors for each particle in the sample reactants and pushes these to the enviroment container.


The observer \textbf{o:}Alignment reads those values, then clears any preexisting pairs and then uses the new values them to calculate the best pairs of eigenvalues to use, based on their alignment (eqn.\ref{eqn:align}).

The observer \textbf{o:}Strength clears the previous strength value then uses this new information to calculate the strength (eqn.\ref{eqn:linkstrength} between the $(i,j)$ pair selected by the alignment, and stores it in the container \textbf{V:}Strengths for use in the next phase.




\subsubsection{Expansion of macro-level \textit{check}()}
The macro-level check() function in the \textbf{a:}link node looks like:
\[
\begin{cases}
pass & r < p_{AB} \And \text{resultant matrix valid}\\
fail & \text{otherwise}
\end{cases}
\]
where $p_{AB}$ is the linking probability  (eqn.\ref{eqn:linkprob}),
and $r$ is a uniformly distributed random value on the interval [0:1].

In the micro-level view we break this down into two decisions. 
The first, \textbf{d:}Prob, decides whether to progress based on  the probability of linking\footnote{%
Here we have only a single probability, for linking two particles.
Full JA-AChem has a more sophisticated approach, allowing multiple particles to link via higher order Jordan products, which involve multiple probabilities.
}, $r < p_{AB}$.




If the probability check passes, 
 we perform some micro-level processing. 
We use a deterministic action (one that always runs in full) \textbf{a:}New\_Mat to calculate what the new matrix would be (eqn.\ref{eqn:jordanprod}) if we did create the new particle. 
We store this for possible later use in \textbf{V:}New\_Mat.

The second decision, \textbf{d:}Valid, uses this result to make its decision based on the trace of the new matrix. If the matrix has a non-zero trace we continue to \textbf{s}:Pull, otherwise we exit the macro-level \textbf{a:}link node without ever pulling information, without destroying existing particles or creating new ones, leaving \textbf{S}:reactant unchanged.


\subsubsection{Expansion of macro-level \textit{pull}()}
This single node action, \textbf{s:}pull, just empties the existing reactants out of the \textbf{S:}reactants container. We delete them by pulling them from the container, so they are no longer accessible.


\subsubsection{Expansion of macro-level \textit{process}()}
This single node, \textbf{a:}New\_Particle, uses the previously calculated matrix stored in \textbf{V}:New\_Mat to create a new particle. 
To do this it creates a link object with the previously calculated strength from \textbf{V:}Strengths.
This link object includes a memory of the reactants that are used to form the link, taken from the local environment container. These properties, the list of eigenstate pairs used, and the matrices are used to generate a new particle object.
The new particle is stored in its own labelled section of the the local storage, \textbf{T:}New\_Part. 

\subsubsection{Expansion of macro-level \textit{push}()}
The sampler \textbf{s:}return moves the newly formed particle in \textbf{T:}New\_Part into the empty \textbf{S:}Reactants.
Control then exits the node. 

\subsubsection{Clearing local containers}
If this functionality is implemented as the macro-level node then the local containers would be deleted when control leaves the node.
In this micro-level implementation we have global containers that persist. We could add an additional observer node just to clear these containers before leaving this section.
In the current implementation we instead give the observers pull access to these containers, and they pull (clear) the variables before they push new values.

\subsection{Modifying JA-AChem}
JA-AChem was first introduced in \cite{Faulkner2016-sz}. 
That JA-AChem system has two properties of its algorithm that we modify here.

\subsubsection{Mass conservation}
The original JA-AChem has no analogue of mass conservation: particles that react are not removed from the system.
So it explores the space of possible composite particles with no limitation of resource, always with a plentiful supply of all discovered particles.
Secondly, the reactions take place in a single well-mixed tank, with no spatial component.

In the classical $(S,R,A)$ model of AChems, any change to the algorithm to change these features is considered a different chemistry (to some degree).  In MetaChem, we can change some features at a macro-level without changing the micro-level detailed linking and reactions: we can change the ``glassware'' without changing the ``chemicals''.

We have mass conservation in the JA-AChem described above.
When a link is formed the components used to make the new particle are pulled from the tank (rather than merely being read) and replaced with the new particle. Similarly when a link decomposes, the components of the link and any remnants of larger particles are processed and returned to the tank.
In this way that the total number of atoms both free and bound within particles remain constant.

\subsubsection{Multiple tanks}
We use MetaChem to introduce multiple tanks, and allow them to interact by transferring particles between the tanks. 
We do this by adding a new outer loop to the overall macro-level flow,
similar to the approach described in StringCatChem (figure~\ref{fig:SCCsystemterm}).

When transferring particles between two tanks we take the contents of both tanks, sort them by size (number of atoms in particle) and then return them to the two tanks by starting with the largest and returning it to a tank and returning the subsequent particles to which ever tank contains less atoms in total. This maintains a rough equality in the number of atoms in each tank.

In section~\ref{sec:nested} we consider single tank, multiple non-interacting tanks, and tanks that interact in a grid or in random organisation. 

There are no transfers when working with a single tank (though we do use a larger single tank with the same number and composition of atoms in the combined contents of the multiple tanks). In multiple tank systems where we choose to have interactions we do this in two ways. In both cases we choose a random number of transfers within a range (in this case 0 to 10 transfers). In the first case of the random transfers we sample 2 tanks without replacement from all the tanks in the system and perform a transfer. In the second (grid transfers) we choose a single tank at random and then select the second tank from the Moore neighbourhood around the first tank.

\section{Swarm Chemistry}
\label{sec:swarm}

We have developed a framework for describing artificial chemistries to replace the limited ($S,R,A$) format. However, all the chemistries we have described in the new framework so far could be built within ($S,R,A$).

In this section we describe Swarm Chemistry \citep{Sayama2009-up}, a system built to explore beyond the ($S,R,A$) format. Despite not having a comfortable description in the ($S,R,A$) framework -- it does not have direct interactions between particles -- SwarmChem is widely known and accepted as an Artificial Chemistry. It is therefore important to show that, while ($S,R,A$) may struggle with SwarmChem, MetaChem comfortably describes it. In the description of SwarmChem in MetaChem we present here, we can see that SwarmChem is not some borderline AChem. It has many close similarities to other more classical AChems when we consider its controls and algorithms, rather than simply its lack of physical connections.

\subsection{Flocking in SwarmChem}

\begin{figure}
    \centering
    \includegraphics{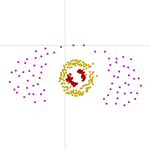}
    \caption[Pulsating Eye swarms contributed to SwarmChem by Benjamin Bush]{Pulsating Eye swarms contributed to SwarmChem by Benjamin Bush using recipe: 102 * (293.86, 17.06, 38.3, 0.81, 0.05, 0.83, 0.2, 0.9)
124 * (226.18, 19.27, 24.57, 0.95, 0.84, 13.09, 0.07, 0.8)
74 * (49.98, 8.44, 4.39, 0.92, 0.14, 96.92, 0.13, 0.51), \url{bingweb.binghamton.edu/~sayama/SwarmChemistry/}. An example of interesting 2D organisation using three different parameter sets for 300 boids.}
    \label{fig:pulsingeye}
\end{figure}

The individuals in SwarmChem, often referred to as boids or agents, interact by each boid changing its own velocity based on the local positions and velocities of its neighbours. This involves no knowledge of the neighbours' internal parameters, just observation of their velocity and position. This gives the effect of swarming or flocking like that seen in birds. Different parameters sets produce different swarms in terms of the density of the swarm and how it moves. In SwarmChem boids with different parameters are allowed to mix (Figure \ref{fig:pulsingeye}).

SwarmChem is a framework for a class of artificial chemistries. Its intention is to explore how higher level statistical rules for chemical systems emerge from lower level local interactions. It does this with the basic concepts of \cite{Reynolds1987-ko}'s Boids. 

\begin{figure}
    \centering
    \includegraphics[width=\linewidth]{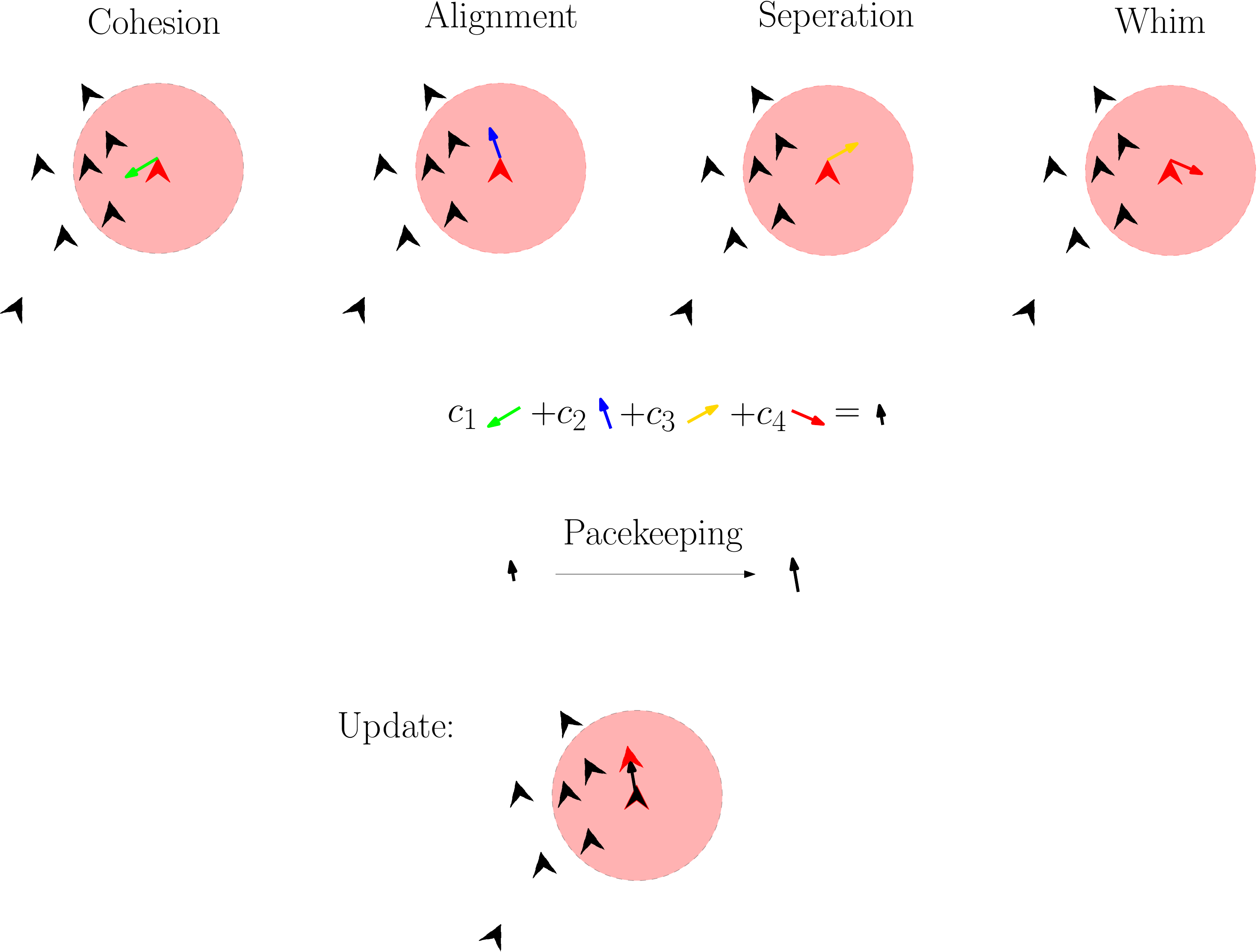}
    \caption[Description of flocking in Reynolds boids]{A pictorial description of flocking in Reynolds boids and swarm chemistry. The red disk shows R the perception distance of our boid.}
    \label{fig:flocking}
\end{figure}

Flocking in both boids and SwarmChem works as follows: at each time step for each boid we first work out the neighbourhood of the boid. We then calculate an acceleration vector of the boid towards the centre of the boids group of neighbouring boids; this is called \textit{cohesion}. We then calculate a vector toward the average heading of the neighbouring boids; this is called \textit{alignment}. We then calculate a vector to prevent crowding, moving to increase the \textit{separation} between boids. Finally we perform ``pacekeeping'', which biases the pace (speed) of the boid towards its normal speed in order to prevent all boid's either becoming stationary or tending towards their maximum speeds.
Then the boid is moved, based on this information. This is done on all boids at once so we use the information of position and velocity from the current time step to calculate the next. See Figure \ref{fig:flocking} for a visual description.

The key change SwarmChem makes to the boids system is to assign \textit{recipes} (parameter sets) to individual boids, rather than using global fixed values. This allows heterogeneous swarms, which can form new kinds of the patterns through their interactions.

In the basic SwarmChem framework,
each boid operates based on a particular recipe; 
extended versions may use multiple recipes with weights used to choose the active recipe \citep{Sayama2010-zp,Sayama2010-zw,Sayama2011-kq}. 
Recipes and weights can be exchanged and changed by other boids. This can be done based on collision or other factors. 

This exchange gives boids a mechanism to change and optimise to maintain structures.
This allows a form of evolution, if we consider a boid to be a child of itself when its parameters change. 

More recently this system has been extended by identifying these larger structures and considering them as entities in their own right \citep{Sayama2018-lj}. Such an approach is important to the analysis of multi-level artificial chemistries.
Initially, the boids were restricted to 2d space, but another extension places the boids in a 3d space \citep{Sayama2012-xu}.

Here we describe a variant of SwarmChem in MetaChem. 
In this variant we exchange a random number of parameters when a collision occurs. This is different from the weighted recipe method used elsewhere. In other versions one boid is dominant in the collision and enforces its recipe on the other, whereas collisions in our system are ``no fault'', in that both boids are changed. We also prevent trading {\tt normal} or {\tt max} parameters if that would result in the boid's  {\tt normal} exceeding its {\tt max}.
A preliminary version of this description can be found in \citep{Rainford2018-xx}.
Here we describe it using a full macro-level graph, and a micro-level graph of the update process, an expansion of the flock action.
We use this description in the nested AChem described in the next section.
This demonstrates that SwarmChem fits comfortably in the MetaChem framework, and can be combined with other AChems.
\subsection{Macro description}

The macro-level graph of our variant of SwarmChem is shown in figure~\ref{fig:swarmmacro}.
It operates as follows:

\begin{figure}
    \centering
    \includegraphics[width=\textwidth]{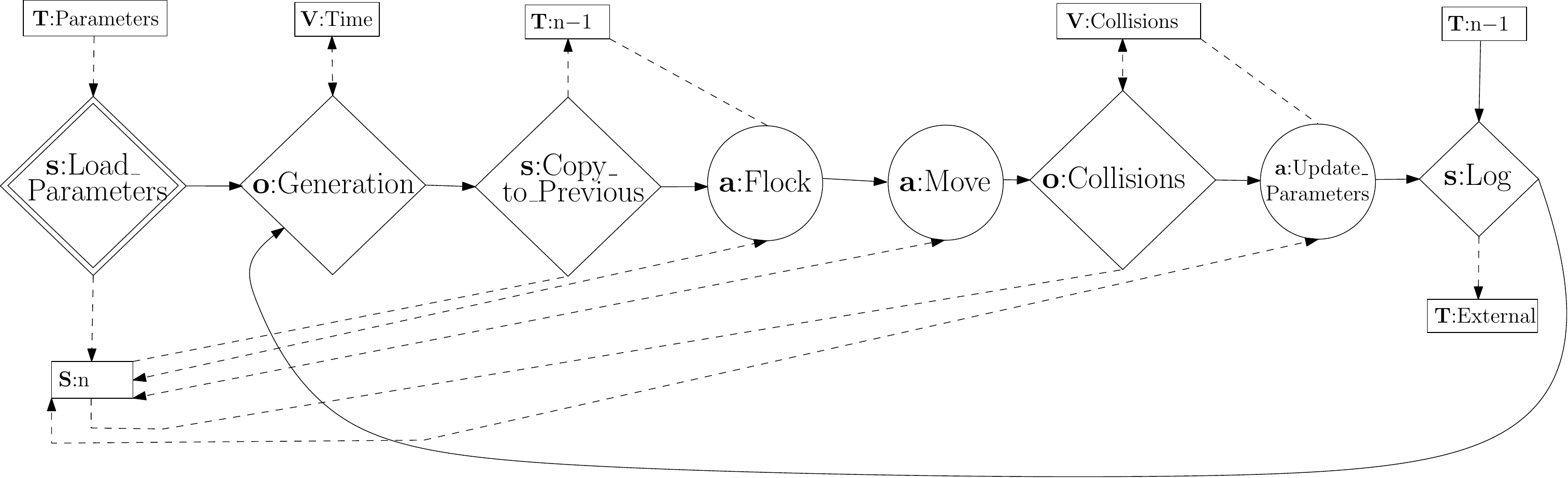}
    \caption[Macro level Swarm Chemistry graph]{Macro level Swarm Chemistry graph. It includes the timing counter \textbf{o:}generation, flocking and moving as well as collisions and the logging sampler to track the chemistry.}
    \label{fig:swarmmacro}
\end{figure}

\begin{itemize}
    \item \textbf{s}:Load Parameters : starting node, which loads the initial parameter set from T:Parameters; and randomly position the boids, stored in \textbf{S}:n.
    \item \textbf{o}:Generation : iterate the clock.
    This ``tick" is part of the discrete timing system that is consistent with all current swarm systems. This is evident in the rest of the macro system as well.
    \item \textbf{s}:Copy\_to\_Previous : the sampler copies the current generation from \textbf{S}:n to the tank \textbf{T}:n-1, which is used to hold the previous generation. This gives a copy of the previous state of all the boids for use in subsequent calculations.
    \item \textbf{a}:Flock : update each boid's parameters (stored in \textbf{S}:n) by following the classic boid rules
    \item \textbf{a}:Move : move all the boids (stored in \textbf{S}:n) based on their parameters and current headings and velocities. This is common to all swarm chemistries.
    \item \textbf{o}:Collisions : check for collisions, and record them in \textbf{V}:Collisions.  This is part of our variant SwarmChem.
    \item \textbf{a}:Update\_Params : update parameter sets that are changed by collision. S:n now contains the fully updated generation.
    \item \textbf{s}:Log : log the previous generation to \textbf{T}:external; clear \textbf{T}:n-1 ready for the current generation to be copied in the next loop iteration
    \item loop back for the next iteration
\end{itemize}

\subsection{Micro description of Flocking}
\label{sec:flock}

\begin{sidewaysfigure}
    \centering
    \includegraphics[width=\textwidth]{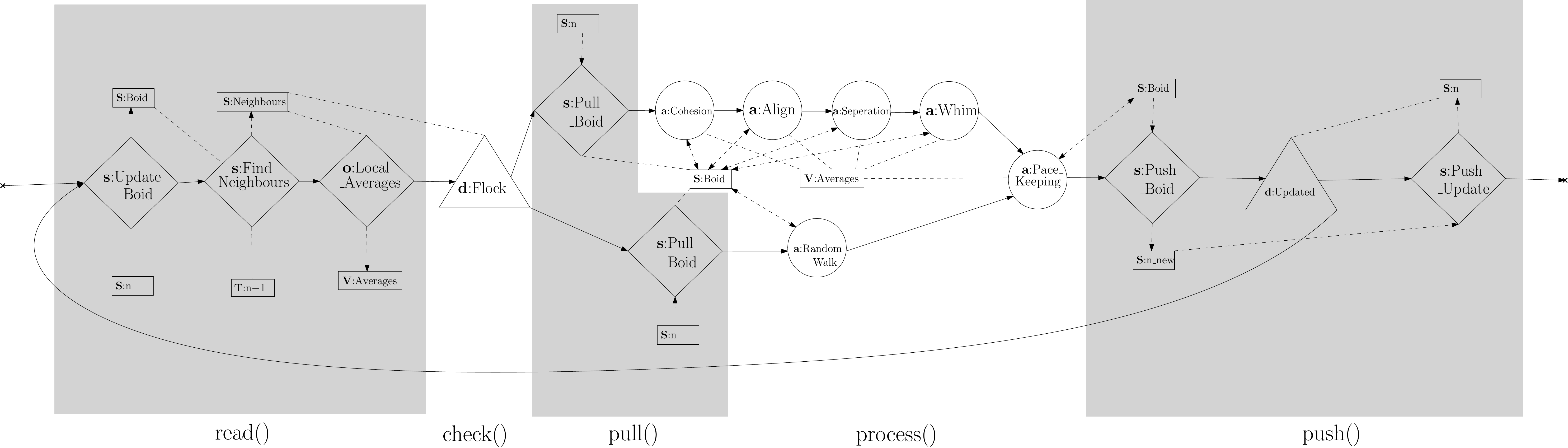}
    \caption[Micro level description of flocking in Swarm Chemistry]{Micro Level description of flocking in Swarm Chemistry. Due to the large loop in this we could consider it a description of many identical nodes or as one node at the Macro level.}
    \label{fig:swarmmicro}
\end{sidewaysfigure}

The flocking action captured in the macro-level \textbf{a}:Flock node (figure \ref{fig:swarmmacro}) contains most of the activity of the system.
In this section we expand that node in a micro-level graph, Figure \ref{fig:swarmmicro}.

As with the JA-AChem example, we do not formally refine each macro-level component function individually, but rather use that structure to guide the design of the micro-level description.
Here we sequentialise the operation, by performing the composition of component functions on each individual particle in the swarm.

\subsubsection{read()}
We start by reading various individuals into different tanks. 
The sampler \textbf{s}:Update\_Boid reads a random single boid from  \textbf{S}:n into \textbf{S}:boid for updating. The sample \textbf{s}:Find\_Neighbours reads out all the neighbours of this boid, defined by its perception distance, into \textbf{S}:Neighbours. 
The observer o:Local\_Averages
 generates $(\Bar{v}),(\Bar{x})$ and $(\Bar{s})$ of the boids in the neighbourhood, and stores them in the environment \textbf{V}:Averages.

\subsubsection{check()}
The action always occurs, so the macro-level check always returns true.
Here we choose to implement a decision to decide the actual process of the function, which is an analogous choice.
The decision node \textbf{d}:Flock 
makes a choice between performing a random walk or normal flocking behaviour, based on the number of neighbours. 
If $|N|>0$ then the operation is flocking, else a random walk.

\subsubsection{pull()}
The sampler s:Pull\_Boid remove the selected boid from the current generation \textbf{S}:n. 
In a subsequent refactoring, we might merge this into the B:Update\_Boid node to simplify the graph.
However, here our initial design is being guided by the transition function format.

\subsubsection{process()}
There are two processing paths.

In the random walk path, a:Random\_Walk sets the current boid with a random velocity (eqn.\ref{eqn:randwalk}).
\begin{equation}
\label{eqn:randwalk}
    \text{Straying:}\quad a_i=(r_{\pm s},r_{\pm s})
\end{equation}

Along the flocking path, four action nodes perform the four calculations of cohesion, alignment, separation, and whim (a small random component added to to motion to keep the system from behaving too predictably) as follows:
\textbf{a}:Cohesion implements eqn.\ref{eqn:cohe},  \textbf{a}:Alignment implements eqn.\ref{eqn:alig},  \textbf{a}:Separation implements eqn.\ref{eqn:sepa}, 
\textbf{a}:Whim implements eqn.\ref{eqn:whim}.
\begin{align}
\label{eqn:cohe}
    \text{Cohesion:}\quad & a_i = c_1(\Bar{x}-x_i) \\
\label{eqn:alig}
    \text{Alignment:}\quad & a_i = a_i + c_2(\Bar{v}-v_i) \\
\label{eqn:sepa}
    \text{Separation:}\quad & a_i = a_i + c_3\Bar{s} \\
\label{eqn:whim}
    \text{Whim:}\quad & a_i = a_i + (r_{\pm s},r_{\pm s})
\end{align}

The branches rejoin at this point, and \textbf{a}:Pacekeeping implements the remaining eqns.\ref{eqn:acce}--\ref{eqn:pace}. The intent here is to prevent boids from constantly increasing in speed, by modifying their speed back towards their normal velocity $v_n$.
\begin{align}\label{eqn:acce}
    \text{Acceleration:}\quad & v_i^* = v_i + a_i \\
\label{eqn:proh}
    \text{Prohibit Overspeeding:}\quad & v_i^* = \text{min}(v_m/|v_i^*|,1)\bullet v_i^* \\
\label{eqn:pace}
    \text{Pacekeeping:}\quad & v_i^* = c_5(v_n/|v_i^*|\bullet v_i^*)+ (1-c_5)v_i^* 
\end{align}

\subsubsection{push()}
Having finished processing, \textbf{s}:Push\_Boid pushes the processed boid to a different sample, \textbf{S}:n\_new, which keeps track of the boids that have been processed. The decision \textbf{d}:Updated decides whether to loop back to process a further boid, or to continue on, based on whether the generation sample \textbf{S}:n has been emptied yet. 
Once all boids have been processed, \textbf{s}:Push\_Update
moves them from \textbf{S}:n\_new back into the (now empty) \textbf{S}:n.

Note that within this graph we do not update velocity; this is done along with position in the macro-level \textbf{a}:Move node.

\subsection{Example: Pulsing Eye}
With this implemented in MetaChem we can still work with the same recipes already generated for Swarm Chemistry. For example here we implemented the recipe shown in Figure \ref{fig:pulsingeye}. In this case we get spinning sets of pulsing eyes which merge into larger and larger eyes. This behaves a little differently because we have a different set of collision rules. But it is visibly the same recipe made of three different parameter sets.

\begin{figure}
    \centering
    \includegraphics[trim = 23mm 0mm 11mm 0.5mm, clip, width=0.7\textwidth]{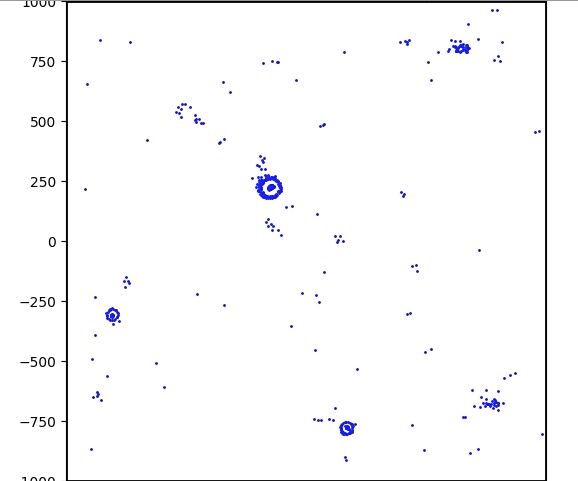}
    \caption{Pulsing Eye recipe as implemented in MetaChem}
    \label{fig:MetaChemPulsinEye}
\end{figure}

\section{Nested Chemistries}
\label{sec:nested}

\subsection{Levels of chemistries}
Sub-symbolic artificial chemistries (ssAChems) \citep{Faulconbridge2009-ch,Faulconbridge2011-pn,Faulkner2018-ex} are generally AChems whose atoms and particles have internal structure that defines their behaviour. 
JA-AChem is one such ssAChem: particles are matrices whose internal structure (elements) defines their linking behaviour through eigenvalues and eigenvectors.
The existing ssAChems are analogous in their rationales to natural chemistry viewed at the level of atomic structure affecting  molecular properties.

Other AChems are designed to reflect the properties of chemistry at the level of cells \citep{Madina2003-rb,Hutton2007-ha} or chemical reaction systems \citep{Soula2016-hw}. In natural chemistry these different levels are closely related: cells contain chemical reaction systems, and chemical reaction systems are based on individual particle and atom interactions. While attempts have been made to bridge the gaps between such levels in individual systems \citep{Liu2018-xu}, so far the systems are very simple and lacking in more complex features.

We can take advantage of feature-rich existing AChems, by using MetaChem to combining them to give a system that can span different levels of activity and behaviour in a single AChem system. 
We demonstrate this approach here by combining JA-AChem \citep{Faulkner2016-sz,Faulkner2017-ap} and SwarmChem \citep{Sayama2009-up,Sayama2010-zw,Sayama2011-kq,Sayama2018-oi}. 

\subsection{General Method}
\begin{figure}
    \centering
    \includegraphics[width=0.4\textwidth]{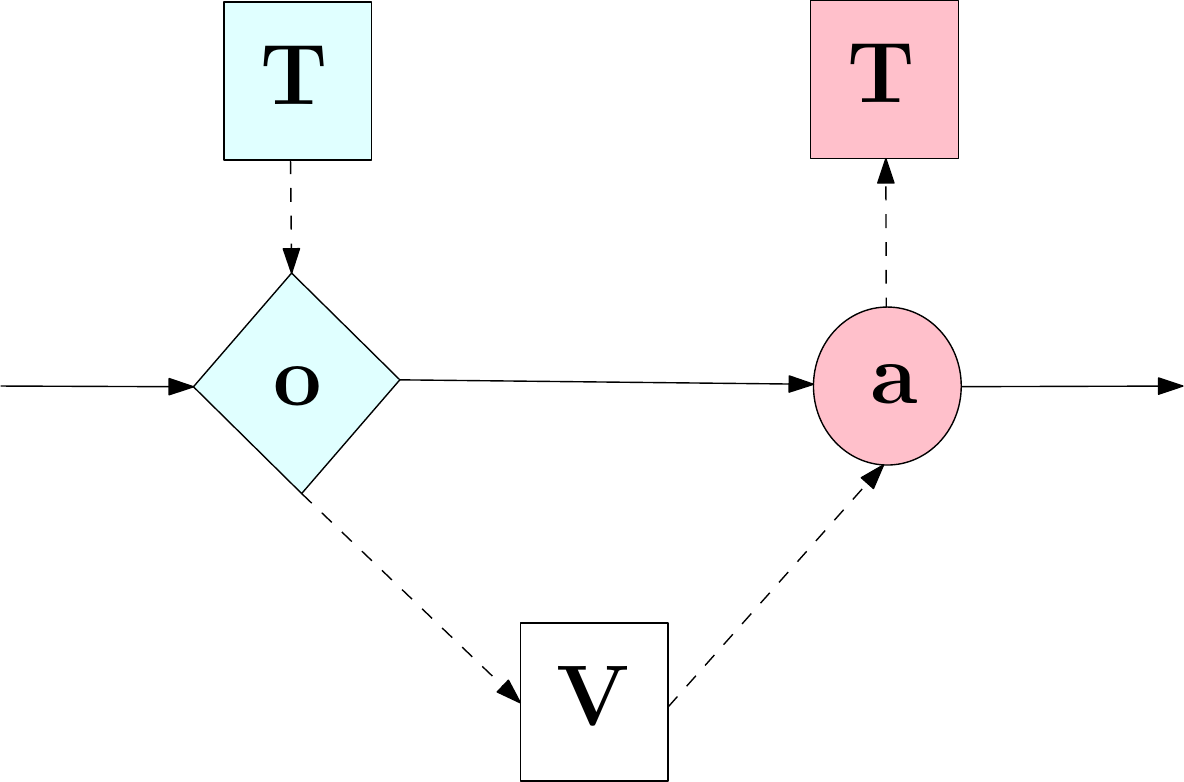}
    \caption[Communication link between two AChems]{Communication link between the blue AChem and the pink AChem. 
    The observer node observes the blue AChem's tank and pushes the communicated information to the shared environment. Control passes to the pink AChem's action node, which acts on the its tank based on information read from the shared environment.}
    \label{fig:comlink}
\end{figure}

We can connect any two AChems in MetaChem by giving them the ability to communicate via their environment \citep{Rainford2018-xx}. This communication can be uni- or bi-directional. 
The basic MetaChem graph structure of combined communicating AChems is given in Figure \ref{fig:comlink}.

We use colour to indicate the `ownership' of a node by a single system. We do not allow a node owned by one AChem to directly communicate with the  nodes owned by a different AChem. 
Instead, information is shared using an environmental container that is not owned by either AChem. 
So in Figure \ref{fig:comlink}, the blue observation is of a tank in the `blue' AChem, and the pink action is on a tank in the `pink' AChem. The figure shows uni-directional communication, in which the blue AChem influences the pink. By adding a second link in the other direction we could establish bi-directional communication. Both the action and the observation are defined by the designer.

For example, if we wish to establish ``side-by-side'' chemistries, where two chemistries with their own separate particles and reactions co-exist in the same spatial system, then our observation will produce a summary statistic that is a value, or set of values, based on the entire system in the first chemistry, which will uniformly affect the entire system in the second chemistry. Alternatively, the observer can generate statistics based on individual particles, which can then affect individual particles in the second AChem. 

Below we give an example of a ``nested'', or multi-level, AChem with bi-directional communication.  The observer of the lower-level AChem generates a set of values over a large number of particles in that AChem. These values are then used to influence the behaviour of a single particle in the higher-level AChem. In turn the behaviour and interaction of one or two particles in the higher-level AChem influence a large number of particles in the lower-level AChem.

\subsection{Implementation: Nested Chemistry}
We generate a new set of chemistries by combining JA-AChem and SwarmChem; 
each of the particles of SwarmChem contains a well-mixed tank of JA-AChem particles (so we have ``swarming tanks of matrix particles''), whose properties inform the SwarmChem particle parameter values.
SwarmChem's spatial movement provides a limitation and control on particle exchanges in JA-AChem between different tanks. The JA-AChem tanks communicate with SwarmChem by changing its parameter values, which influences the agents' spatial movement and likelihood of collision.

First, we abstract the description of both AChems to a higher level that comprises two control nodes and two container nodes.
The first control node, \textbf{s}:LoadX, is the initialisation sampler, that loads the initial state from \textbf{T}:InitX into \textbf{T}:X.
The other control node, a:UpdateX, performs one of the outer loops of the AChem's operation as defined in the earlier macro-level graphs.

The other control nodes associated with each of our separate chemistries is new and deals with modifying the associated chemistry based on information observed from the other chemistries particles.

\begin{figure}
    \centering
    \includegraphics[width=\textwidth]{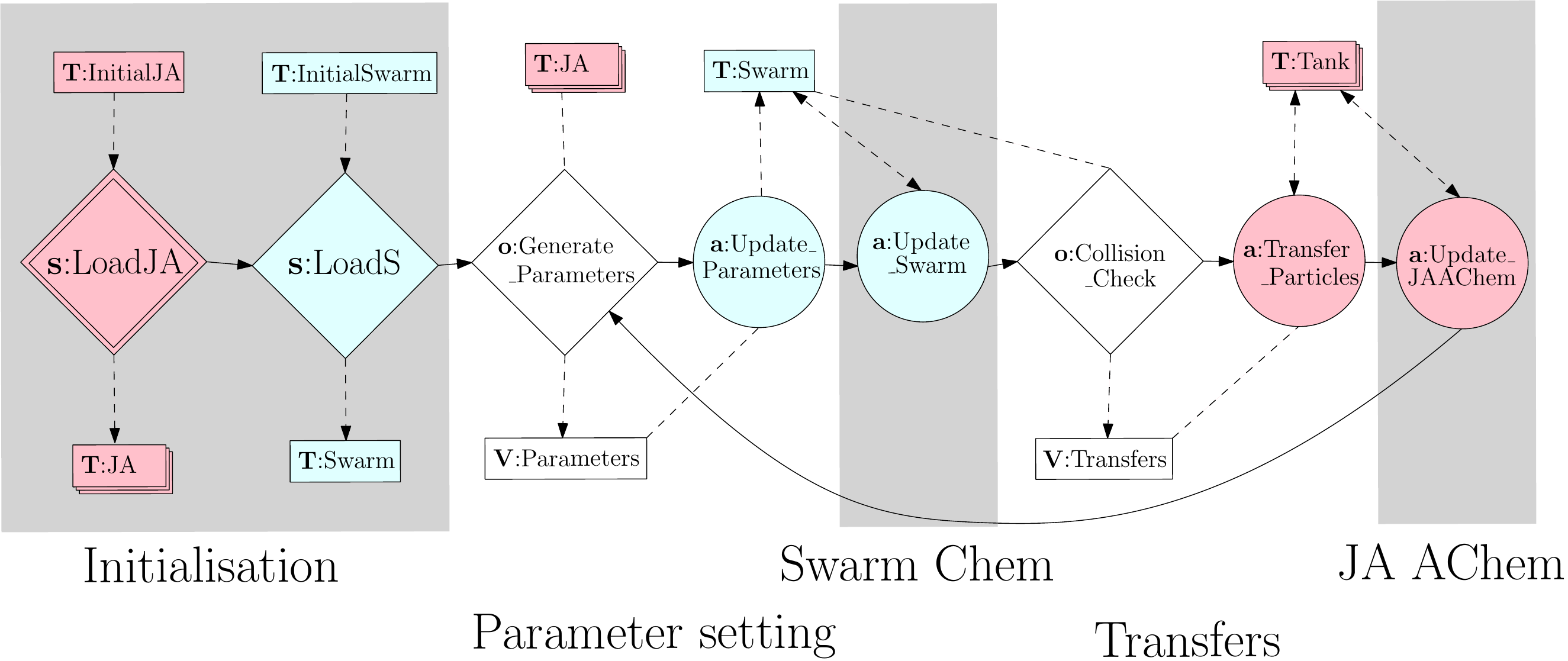}
    \caption[Macro-level NestedChem graph]{Macro-level NestedChem  graph in MetaChem. JA-AChem nodes are shown in pink, SwarmChem nodes in blue. White nodes are either shared or not natively part of either AChem.}
    \label{fig:system}
\end{figure}

We link these individual high-level AChem graphs in various ways to give seven distinctive systems; an eighth system is achieved through a change in system settings.
The largest of these systems is a fully nested AChem that contains all the AChem and linking nodes used in our systems, Figure \ref{fig:system}.

We combine the systems with two graph fragments, labelled Parameter Setting and Transfers in Figure \ref{fig:system}.
These provides a means of communication between the two systems.  
The five stages shown in Figure~\ref{fig:system} are:
\begin{description}
\item[Initialisation:] Initial tanks of JA-AChem particles and initial swarm agents are loaded into the system and stored separately with matching indexing to allow for reference between the two.
\item[Parameter setting:]  This generate parameters values for each SwarmChem agent based on the particles in its associated JA-AChem tank. 
The parameter values are pushed to the environmental container \textbf{V:}parameters (figure \ref{fig:system}). The swarm then updates itself by reading these values.
\item[SwarmChem:] The SwarmChem particles are updated and moved using a single SwarmChem timestep.
\item[Transfer:] SwarmChem assesses whether any collisions have occurred between its particles. It pushes a record of these collisions to the environmental container \textbf{V:}transfers. JA-AChem reads this container, and uses the results to exchange particles between tanks based on the SwarmChem collisions.  

\item[JA-AChem:] The JA-AChem updates by performing a number
of  bonding and decomposition attempts. All tanks are independent  mass-con\-serving well-mixed tanks.
\end{description}

There are apparently four invalid edges in the macro system graph of NestedChem (figure~\ref{fig:system}): (\textbf{a:}Update\_Parameters, \textbf{T:}Swarm), (\textbf{a:}Swarm\_Update, \textbf{T:}Swarm), (\textbf{a:}Transfer\_Particles, \textbf{T:}Tank) and (\textbf{a:}JA-AChem\_Update, \textbf{T:}Tank). All of these edges appear to allow actions to push to tanks, which is not allowed (table~\ref{tab:infolimits}). In the case of \textbf{a:}Swarm\_Update and \textbf{a:}JA-AChem\_Update we have seen the expanded graphs of these nodes in the macro graphs of each system, Figures \ref{fig:JAMacro} and \ref{fig:swarmmacro}. In those graphs we see that the actions carried out by these nodes mean they always move the particles to samples before making any changes. Here we connect directly to the tanks as this is a macro graph, a form of pseudo-code, this is actually implemented with these nodes expanded through the macro graphs shown previously down to the micro graph, Figures~\ref{fig:JAACalink} and~\ref{fig:swarmmicro}. At these levels of description the content of these tanks is moved to samples before being used.

In the case of \textbf{a:}Transfer\_Particles, if we were to expand this node we would see that all the operations of this node are carried out by samplers, and there is therefore no issue that before starting the particles have not been moved to a sample. 

Finally, in the case of \textbf{a:}Update\_Parameters the process function is applied over all particles in the system, meaning the sample would be the entire tank, so in another abuse of notation and to avoid introducing a further two control nodes and a container to move the entire contents back and forth, we allow the node to connect directly to the tank. It should be take given that in the expanded form the necessary sampling would occur.

\subsubsection{Modular Systems}
\begin{figure}[t]
\centering
\adjustbox{valign=t, max width=\textwidth}{\begin{tikzpicture}
\draw [fill=gray!10, gray!20] (-0.5,0.5) rectangle (2,-12.5);
\draw [fill=gray, gray!20] (4.75,0.5) rectangle (6.25,-12.5);
\draw [fill=gray, gray!20] (9,0.5) rectangle (10.5,-12.5);
\node (C1a) at (0,0) [shape=diamond,fill=gray!90,draw] {};
\node (Name) [left=of C1a] {\textbf{I}:};
\node [shape=diamond,fill=gray!30, draw] (C1b) [right=of C1a] {};
\node [shape=diamond, fill=white,draw] (C2a) [right=of C1b] {};
\node [shape=circle, fill=gray!30,draw] (C2b) [right=of C2a] {};
\node [shape=circle, fill=gray!30,draw] (C3) [right=of C2b] {};
\node [shape=diamond, fill=white,draw] (C4a) [right=of C3] {};
\node [shape=circle, fill=gray!90,draw] (C4b) [right=of C4a] {};
\node [shape=circle, fill=gray!90,draw] (C5) [right=of C4b] {};
\draw[-triangle 45](C1a) -- (C1b);
\draw[-triangle 45] (C1b) -- (C2a);
\draw[-triangle 45] (C2a) -- (C2b);
\draw[-triangle 45] (C2b) -- (C3);
\draw[-triangle 45] (C3) -- (C4a);
\draw[-triangle 45] (C4a) -- (C4b);
\draw[-triangle 45](C4b) -- (C5);
\draw[-triangle 45] (C5.south east) to [bend left=10] (C2a.south east);
\node (C1a) at (0,-2) [shape=diamond,fill=gray!90,draw] {};
\node (Name) [left=of C1a] {\textbf{II}:};
\node [shape=diamond,fill=gray!30, draw] (C1b) [right=of C1a] {};
\node [shape=diamond, fill=white,draw] (C2a) [right=of C1b] {};
\node [shape=circle, fill=gray!30,draw] (C2b) [right=of C2a] {};
\node [shape=circle, fill=gray!30,draw] (C3) [right=of C2b] {};
\node (C5) at (9.7,-2) [shape=circle, fill=gray!90,draw] {};
\draw[-triangle 45](C1a) -- (C1b);
\draw[-triangle 45] (C1b) -- (C2a);
\draw[-triangle 45] (C2a) -- (C2b);
\draw[-triangle 45] (C2b) -- (C3);
\draw[-triangle 45] (C3) -- (C5);
\draw[-triangle 45] (C5.south east) to [bend left=10] (C2a.south east);
\node (C1a) at (0,-4) [shape=diamond,fill=gray,draw] {};
\node (Name) [left=of C1a] {\textbf{III}:};
\node [shape=diamond,fill=gray!30, draw] (C1b) [right=of C1a] {};
\node (C2b) at (4.2,-4) [shape=circle, fill=gray!30,draw] {};
\node [shape=circle, fill=gray!30,draw] (C3) [right=of C2b] {};
\node [shape=diamond,fill=white, draw] (C4a) [right=of C3] {};
\draw[-triangle 45](C1a) -- (C1b);
\draw[-triangle 45] (C1b) -- (C2b);
\draw[-triangle 45] (C2b) -- (C3);
\draw[-triangle 45] (C3) -- (C4a);
\draw[-triangle 45] (C4a.south east) to [bend left=15] (C2b.south east);
\node (C1a) at (0,-6) [shape=diamond,fill=gray!90,draw] {};
\node (Name) [left=of C1a] {\textbf{IV}:};
\node [shape=diamond,fill=gray!30, draw] (C1b) [right=of C1a] {};
\node (C3) at (5.5,-6) [shape=circle,fill=gray!30, draw] {};
\draw[-triangle 45](C1a) -- (C1b);
\draw[-triangle 45] (C1b) -- (C3);
\draw[-triangle 45] (5.6,-5.8) arc (150:-150:10pt);
\node (C1a) at (0,-8) [shape=diamond,fill=gray!90,draw] {};
\node (Name) [left=of C1a] {\textbf{V} \& \textbf{VI}:};
\node [shape=diamond,fill=gray!30, draw] (C1b) [right=of C1a] {};
\node (C5) at (9.7,-8) [shape=circle,fill=gray!90, draw] {};
\draw[-triangle 45](C1a) -- (C1b);
\draw[-triangle 45] (C1b) -- (C5);
\draw[-triangle 45] (9.8,-7.8) arc (150:-150:10pt);
\node (C1a) at (0,-10) [shape=diamond,fill=gray!90,draw] {};
\node (Name) [left=of C1a] {\textbf{VII}:};
\node [shape=diamond,fill=gray!30, draw] (C1b) [right=of C1a] {};
\node (C5) at (9.7,-10) [shape=circle,fill=gray!90, draw] {};
\node [shape=circle, fill=gray!90, draw] (C4b) [left=of C5] {};
\node [shape=diamond, fill=white, draw] (C4a) [left=of C4b] {};
\draw[-triangle 45](C1a) -- (C1b);
\draw[-triangle 45] (C1b) -- (C4a);
\draw[-triangle 45] (C4a) -- (C4b);
\draw[-triangle 45] (C4b) -- (C5);
\draw[-triangle 45] (C5.south east) to [bend left=15] (C4a.south east);
\node (C1a) at (0,-12) [shape=diamond,fill=gray!90,draw] {};
\node (Name) [left=of C1a] {\textbf{VIII}:};
6\node [shape=diamond,fill=gray!30, draw] (C1b) [right=of C1a] {};
\node (C5) at (9.7,-12) [shape=circle,fill=gray!90, draw] {};
\node [shape=circle, fill=gray!90, draw] (C4b) [left=of C5] {};
\draw[-triangle 45](C1a) -- (C1b);
\draw[-triangle 45] (C1b) -- (C4b);
\draw[-triangle 45] (C4b) -- (C5);
\draw[-triangle 45] (C5.south east) to [bend left=20] (C4b.south east);
\end{tikzpicture}}
\caption{Various combinations of JA-AChem and SwarmChem, See text for details.}
\label{fig:combinations}
\end{figure}
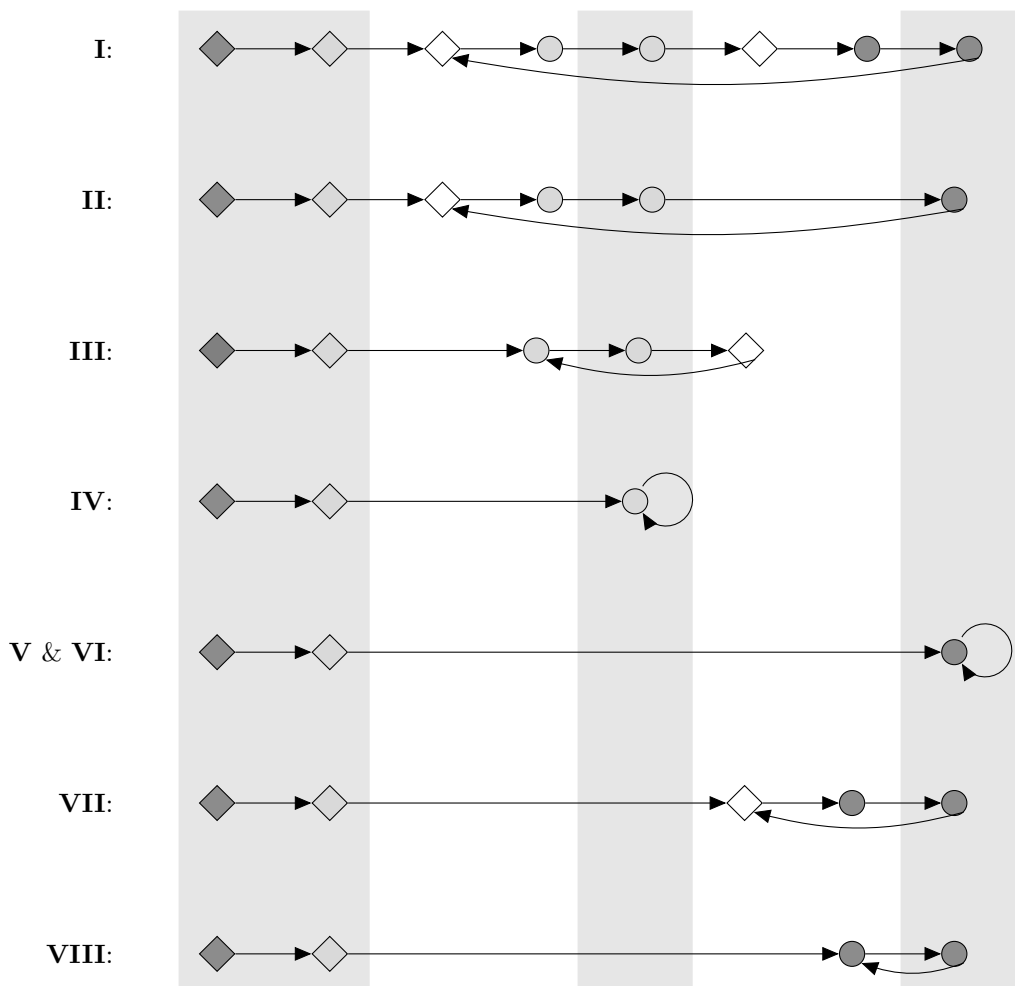

From this full system we can derive eight variant systems. The control flow of these systems is shown in Figure~\ref{fig:combinations}. 


\begin{description}
\item[I. Nested.] The full Nested AChem system as shown in Figure \ref{fig:system}
\item[II. Nested without collision.] JA-AChem particles are not transferred between tanks, but still determine the parameter values of agents in the SwarmChem
\item[III. SwarmChem.] SwarmChem agents randomly exchange parameter values on collision; there is no communication with the JA-AChem.
\item[IV. SwarmChem without collision.] A very basic form of SwarmChem in which the agents  interact only through Boid like flocking behaviours.
\item[V. JA-AChem single tank.] A single well-mixed tank of JA-AChem. The same number of evaluations are used per generation and the same number of starting particles are also used as the other systems.
\item[VI. JA-AChem multiple tanks with no interaction.] A JA-AChem  with the same number of tanks as in the nested version; there are fewer atoms and particles in each tank, but the same number of overall atoms and evaluations are used.
\item[VII. JA-AChem multiple tanks with random transfers.] The same system as in \textbf{VI} but with tanks randomly selected to randomly transfer particles between them.
\item[VIII. JA-AChem multiple tanks with grid transfers.] The same as in \textbf{VII} but transfer tanks selected based on a Moore Neighbourhood
\end{description}


\subsubsection{Discussion}

In the JA-AChem level of the systems the resultant number of particles in the tanks should quickly stabilise, but we expect the systems with transfers to be less stable than others. Particles being transferred in and out of the tanks should disturb any equilibrium.

We also expect to see larger particles in the partitioned systems as the smaller size of the tanks limit the sampling possibilities, increasing the chances of selecting molecules which already contain multiple particles. As these are used and the number of particles in the tank decreases, these probabilities should further increase.

We can observe many different statistics on the agents of the swarm. In homogeneous flocking the relative position of an agent to its visible neighbours should be very similar across agents, as a flock all have the same perception radius and tendency for avoidance. In SwarmChem these have greater variation but should be similar in sets of agents forming a swarm. Here we expect to see greater variation in the nested SwarmChem where all values of perception radius and tendency for avoidance are possible.

Results, analysis and further discussion of these nested systems can be found in \citep{Rainford2018-xx,Rainford2018}.

\section{Conclusion and Future Work}
\subsection{Conclusion}
We now have a formal language in which to discuss different AChems. 
In MetaChem, we can readily take the same ``chemicals'', and investigate their behaviour in different ``glassware'',
separating out the contribution of the underlying low level bonding rules from the environmental effects of how the particles are brought together.

All current systems we are aware of in the literature can be described by static graph MetaChem. 

To show the power of this modularisation and graph-based representation, we have presented two case studies of AChems. The first of these is our own chemistry JA-AChem, originally developed based on algebraic structures, but here re-described in the MetaChem format. The second AChem is Swarm Chemistry, chosen for being a well-established AChem that is not well-described in the $(S,R,A)$ format. SwarmChem and JA-AChem are very different AChems, with next to no overlap in their nature. SwarmChem also represents an independent example of description of an existing AChem in MetaChem.

We combine these using a graph structure that we widely applicable for the joining of two artificial chemistries. This is one possible use of the Static Graph MetaChem, there are others and extensions of MetaChem described below.

\subsection{Future Work}

\subsubsection{Other Combinations of AChems}
We have illustrated one approach for combining artificial chemistries here, with two specific AChems, but the potential is much broader. 
New and different hierarchical AChem combinations could be tried. 
There are also mixed systems and joint systems to consider. 
Mixed systems would share tanks or spatial environment, with a new interaction added between the different types of particles. 
Joint systems would work with a combined particle made up of a particle from each system.

\subsubsection{Static Graph MetaChem: reuse and toolsets}
The \textit{static graph MetaChem} described here is a first step towards standardised framework for AChems. It is not just a mathematical framework; it can and has been implemented in software \citep{Rainford2019-ALife}. These graphs provide more than a simple visualisation: they are a new way to design, implement and run AChems.

MetaChem provides a move forward in designing AChems. As we gather more descriptions in this framework, designers can begin make use of parts of existing descriptions in new systems. 
Additionally, we can start to standardise output values from AChems, and make use of standard visualisations and reporting of results. 
The use of modularised structured nodes with defined functions should allow designers to define new nodes easily.

We have defined a general method of composition using indirect communication (a form of environment orientation \citep{Hoverd2009-vo}) with macro-level graphs. The ability to join systems and use modularity to share parts of algorithms could provide, after more development, significant speedups in designing and implementing new AChems.

The modularity and clear designation of particles and environmental properties allows the design of generic analysis tools, visualisation tools, and metrics that can be used across similar systems. 
An AChem can provide a set of particles and their position to a visualiser, regardless of the system's other properties.
A more general purpose proximity-based analysis for higher level object identification, such as that used in more recent SwarmChem work \citep{Sayama2018-zz, Sayama2018-lj}, becomes reusable.

\subsubsection{Dynamic MetaChem}
\label{sec:dynamic}
Above we describe \textit{static graph} MetaChem. The graph exists before the system is run and does not change at run time, similar to most programs. 
A static graph MetaChem could be defined with a set of graph-rewriting rules that generate the graph. The rule set could  produce a particular graph or multiple possible graphs, such as the ones in figure~\ref{fig:combinations}. 

This use of graph-rewriting rules then provides a natural way to make the topology dynamic, during AChem execution.
A system could grow at run time, and could grow differently dependent on differences in the produced particles and variables.

A \textit{dynamic edge graph} MetaChem would allow the graph to add and remove edges during run time. 
This could use a further type of control node to be responsible for this rewrite. In terms of the hierarchy, Figure \ref{fig:structure}, these nodes would fall under the grouping of control flow admin nodes. 
Such a system could reorder its own control flow, or even connect entirely new nodes or subgraphs to the control flow that, while existing at the start, were not connected.

We could use these new control nodes to trigger events in the system based on specific conditions, such as complexity. We could also use this as part of evolving Artificial Chemistries, by having a set of nodes to start with and allowing edges to change over time until the control flow becomes stable. This could be controlled by the system itself, so it would ``learn" an artificial chemistry.

\subsubsection{Evolving MetaChem}

For true evolution and change we would move to full \textit{graph language} MetaChem (or \textit{dynamic graph} MetaChem), which could create and destroy its own nodes and edges at run time.
The graph could grow, and could remove parts that were no longer needed, allowing it to prune its own process. This type of system would allow the AChem to change completely at run time, so it could truly transition and change abstraction levels and experiments as it ran. This could enable paths in open-ended evolution and open-ended systems research.

If we 
allow the set of particles to be the graph \textit{rules} that generate the graph, and the reactions change those rules, then we can evolve how graphs form. 
This would allow systems not only capable of changing at run time but of changing their basic components and how they can run during their execution. With suitable initial rule sets and reactions this could allow for the production of a completely unexpected AChems with as little design bias as possible.

This could allow the design and growth of a system capable of self-reflection and change at run time. This opens new possibilities  for transitions towards open-ended evolution \citep{Stepney:2011:reflect}, as we can build systems capable of reacting to new emergent objects or behaviours if they can be identified. For example, if a system identified a set of objects within itself, it could then attempt to model those objects at a higher level and improve that model with information from the original low-level implementation \citep{Nellis:2010:levels}.

\subsubsection{MetaChem as an AChem}
We can consider MetaChem as an AChem itself, where the atoms  are graph nodes, the links are graph edges, and the composite particles are (potental) AChems. 
We can consider an isolated MetaChem subgraph as forming a subAChem or a full AChem. 
An instance of MetaChem is therefore not a single graph but a collection of graphs. So MetaChem provides both the language to describe AChems, and a process to build, compose, and evolve AChems.  This is a key advantage of using a graph-based formal framwork in the definition of MetaChem.


\section{Acknowledgements}
The research for this work was done with PhD funding from the Department of Chemistry, University of York, UK.

\bibliography{Rainford}

\appendix

\section{Appendix:  Mathematical Formalism}
\label{sec:maths}
We provide a mathematical formalism here.
First we describe the static elements that make up the static graphs of our system. Then we define the dynamic system state.
Finally, we define the state transition function over the graphs that are used to capture the dynamics of a specific AChem.

\subsection{Static Graph MetaChem}
We have two static graphs capturing the control flow and information flow.  We build up the definition as follows.

\subsubsection{Nodes}
The set of graph nodes is $N$.
As shown in our hierarchy, Figure \ref{fig:structure}, our system is composed of container and control nodes.  
The graph nodes are partitioned into two sets: Control nodes $C$ and Container nodes $B$.  
\begin{equation}
    \langle B,C \rangle ~\mathbf{\sf partition}~ N
\end{equation}
where the notation $\langle X_1, \ldots, X_n \rangle~\mathbf{\sf partition}~X$  means that the set $X$ is partitioned by the $n$ subsets $X_i$. 

Each of the  sets $B$ and $C$ is  partitioned further. 

The container nodes $B$ comprise three categories of node: Environment nodes $V$, Tank nodes $T$, and Sample nodes $S$. 
\begin{equation}
\langle V,T,S\rangle ~\mathbf{\sf partition}~ B
\end{equation}

Control nodes are more complicated in the hierarchy but the static components partition the set into Action ($C_a$), Decision ($C_d$), Observer ($C_o$), Sampler ($C_s$), and Termination ($C_t$) nodes.
\begin{equation}
\langle C_a,C_d,C_o,C_s,C_t \rangle ~\mathbf{\sf partition}~ C
\end{equation}

\subsubsection{Edges}
Our hierarchy, Figure \ref{fig:structure}, also contains control flow and information flow. These appear in our static graphs as edges.
We define an edge as a pair of nodes.
Edges are either control edges or information edges, $E_G$ and $E_I$.

\paragraph{Control Edges.}
These are edges between control nodes
\begin{equation}
E_G\subseteq C\times C  
\end{equation}

Different subtypes of control nodes can have different numbers of exiting edges.
Define $\mbox{\bf \sf target}$
to map a source control node to the set of target control nodes connected to it by an edge in $E_G$:
\begin{align}
    &\mbox{\bf \sf target} : C \rightarrow {\mathbb P} C \\
    &\forall c: C \,|\, \mbox{\bf \sf target}(c) = \{ c_s : C \,|\, (c,c_t) \in E_G \}
\end{align}

Decision control nodes have multiple targets;
all other control nodes have a unique target:
\begin{align}
    & \forall c:C_d \,|\, \# \mbox{\bf \sf target}(c) > 1 \\ 
    & \forall c: C \setminus C_d  \,|\, \# \mbox{\bf \sf target}(c) = 1 
\end{align}

\paragraph{Information Edges.}
These come in three varieties.

Read edges, $E_{read}$, are directed from control nodes to containers, and indicate which containers' information a control node can read. In the graphical notation they are shown as undirected edges, as there is no change to the container node. 
\begin{equation}
E_{read} \subseteq C\times B
\end{equation}

Pull edges, $E_{pull}$, are directed from containers  to control  nodes, and indicate the containers that a control node can remove information or objects from.
Every $E_{pull}$ edge must have a corresponding $E_{read}$ edge:
\begin{equation}
E_{pull}\subseteq B\times C, \quad E_{pull}\subseteq E_{read}^{-1}
\end{equation}

Push edges, $E_{push}$, are directed from control nodes to containers, and indicate the containers that a control node can push information and objects to. 
Every $E_{push}$ edge must have a corresponding $E_{read}$ edge.
\begin{equation}
E_{push}\subseteq C\times B, \quad E_{push}\subseteq E_{read}
\end{equation}

There are limits on the containers that different node types are allowed to have pull and push edges with; see Table \ref{tab:pullpush}.
\begin{table}[t]
    \centering
    \begin{tabular}{lccccc}\toprule
        & action & decision & sampler & observer & termination\\
        \midrule
        tank $T$ & & & \checkmark & \\
        sample $S$ & \checkmark & & \checkmark & \\
        variable $V$ & \checkmark & & & \checkmark \\
        \bottomrule
    \end{tabular}
    \caption{Container types that Control nodes are allowed to have push and pull edges with.}
    \label{tab:pullpush}
\end{table}
\subsubsection{Graphs}

We have two graphs, $G_\Omega$, $I_\Omega$, of our system $\Omega$, capturing control and information  respectively. For the purpose of this definition we  assume we are always referring to elements of a given system, and so  drop the system label such that our graphs $G_\Omega$, $I_\Omega$ become $G$ and $I$.

\begin{equation}
G = (N,E_G),\qquad I = (N,E_I)
\end{equation}

\subsection{Dynamic System State}
\label{app:maths:state}
 
Now we have these static graphs, we can start a dynamic process guided by them. 
We denote the dynamic aspects of our system using the Greek alphabet, to distinguish it from static components.

Container nodes $B$ can contain particles of type $\Phi$.
The structure of the set $\Phi$ of particle types is application dependent.
We define the contents $P$ of such a node as a bag (multiset) of particle types:
\begin{equation}
    P = \Phi\rightarrow \mathbb{N}
\end{equation}
where $\mathbb{N}$ is the set of natural numbers  counting how many instances of each particle type there are in the bag.
Here we refer to the content of a container using the mapping from container to particle bag; in the informal sections above we abuse the notation and refer to the content of containers simply by the container label, for readability and brevity.

Environment nodes $V$ contain environment information of type $\Psi$.
We do not here further define the structure of the set $\Psi$; this is application dependent.

The current state node $c \in C$,
a pointer in this static graph case, is a control node dynamically assigned and changing over time. This pointer indicates the current control node, whose transition function is to be run in order to find the next state of the system. 

The full system state comprises five components: the Control Graph $G$, the Information Graph $I$, the current state node $c \in C$, a mapping from the container nodes to the bag of particles they contain ($\phi \in B\rightarrow P$), and a mapping from the environment nodes to the dynamic environment information they contain ($v \in V\rightarrow\Psi$).  
The system being formalised here has static graphs $G$ and $I$, so we do not here include them in the system state, rather taking them to be globally defined.
The set of system states is: 
\begin{equation}
\Omega_s = C\times (B\rightarrow P)\times (V\rightarrow \Psi)
\end{equation}

We define a specific system state $\omega$ as the triple $(c,\phi,\psi) \in \Omega_s$. 
The initial state of the system has $c=c_0$, the identified start node.

\subsection{Transition Functions}
Each node has a transition function of the whole system state, as nodes can access and affect neighbouring nodes\footnote{%
It would be possible to define local state transition functions, and ``promote'' them \citep{SS-ZB-03b} to a global state transition function, but the mathematical machinery needed to do so is here more cumbersome than a direct definition.
}.
\begin{equation}
\delta: \Omega_s \rightarrow \Omega_s
\end{equation}

The overall transition function is decomposed into the component functions: $read()$, $check()$, $pull()$, $process()$  $push()$, $next()$. For any node some of these may be null (identity) functions. 
For some kinds of nodes, some components are always null or defaulted, Table \ref{tab:transitionfunction}.

Using $\fatsemi$ to indicate strict ordering of application of functions from left to right gives the following definition of $\delta$:
\begin{equation}
\delta = pull \fatsemi process \fatsemi push \fatsemi check \fatsemi read \fatsemi next
\end{equation}

Each of these transition function components plays a different role in the transition and uses a different aspect of the state.
These functions are summarised in Figure~\ref{fig:transsum} and formalised below.

The transition function components exploit a local state, which exists only for the duration of the transition. 
This comprises a labelled bags of particles (labelled by their source container node), and labelled environment variables (labelled by their source environment node)
\begin{equation}
Local = (B\rightarrow P)\times (V\rightarrow \Psi)
\label{eqn:localstate}
\end{equation}
We define a specific local state as the pair $(\phi_l,v_l) \in Local$. 

Local state disappears as soon as the transition function is completed, so control nodes have no lasting state or memory. Any information used by a control node must come from containers at the start of a transition using the $read()$ or $pull()$ functions, and any information or objects that should remain in the system should be written back to a containers by the $push()$ function.

\subsubsection{read()}
The $read()$ function allows a node to collect information from external containers into the temporary local containers, where it can to be used by the following transition functions. This does not modify the system state. 
\begin{align}
    &read : \Omega_s \rightarrow (\Omega_s \times Local)\\
    &read(c,\phi,\psi) = ((c,\phi,\psi), (\phi_l,\psi_l))
\end{align}

The containers and environment nodes from which current node $c$ can read are the ones attached by read edges:
\begin{align}
    B_c = \{ b \in B \,|\, (c,b) \in E_{read}\} \\
    V_c = \{ v \in V \,|\, (c,v) \in E_{read}\}
\end{align}

The default behaviour of $read()$ is to copy all the readable containers to the local state: 
\begin{align}
    \phi_l = \{ (b,\rho) \,|\, b \in B_c \land (b,\rho) \in \phi \} \\
    \psi_l = \{ (v,p) \,|\, v \in V_c \land (v,p) \in \psi \}
\end{align}
In practice, implementations may choose to read only a subset of the information in the readable containers.

\subsubsection{check()}
The $check()$ function uses the local information to generate a threshold probability, which is used to determine whether the rest of the transition (the part that actually alters containers) occurs, or exits at this point.
This packages any  probabilistic aspects of the execution of the transition.

The check function uses a probability spawning function (psf) \citep{Faulkner2017-ap} to determine if the rest of the transition will occur. The default behaviour in this case is to return True, which it does for administrative nodes which always operate in a deterministic manner.

The execution checks the generated probability $\mbox{\it psf}(\phi_l,\psi_l)$ against a uniform random number, $r$.
If our threshold probability is less than $r$ we continue, otherwise we exit.
\begin{equation}
\begin{cases}
\delta = next & ; \mbox{\it psf}(\phi_l,\psi_l) < r\\
check = Id(\Omega_s \times Local)  & ; \mbox{otherwise} 
\end{cases}
\end{equation}
where $r\in[0:1]$ is a uniformly distributed random number. 

Either the transition function exits and does not proceed to any further functions, else the other functions are executed as expected.
In either case, $check$ makes no change to the state of the system, or the local state. 
\subsubsection{pull()}
The $pull()$ function removes information from connected containers. Any information removed has (potentially) been copied to the local state in $read()$, where it is available for local processing. 
This modifies the system state, but not the local state
\begin{align}
    &pull : (\Omega_s \times Local) \rightarrow (\Omega_s \times Local)\\
    &pull((c,\phi,\psi), (\phi_l,\psi_l)) = ((c,\phi',\psi'), (\phi_l,\psi_l))
\end{align}

The containers and environment nodes from which current node $c$ can pull are the ones attached by pull edges:
\begin{align}
    B_c = \{ b \in B \,|\, (b,c) \in E_{pull}\} \\
    V_c = \{ v \in V \,|\, (v,c) \in E_{pull}\}
\end{align}

The pull function may only change containers connected by pull edges, and then only to delete information from them\footnote{%
$\mathbf{\sf subbag}$ has the obvious definition: there may not be more particles of any given type after than before.
$\mathbf{\sf subenv}$ is application dependent, but should conform to the idea of removing information.
}:
\begin{align}
    \forall b \in B_c \,|\, \phi'(b)  ~\mathbf{\sf subbag}~ \phi(b) \\
    \forall b \in B \setminus B_c \,|\, \phi'(b)  = \phi(b) \\
    \forall v \in V_c \,|\, \psi'(v)  ~\mathbf{\sf subenv}~ \psi(v) \\
    \forall v \in V \setminus V_c \,|\, \psi'(v) = \psi(v) 
\end{align}
The default behaviour of pull is to do nothing: $pull = Id(\Omega_s \times Local)$. 


\subsubsection{process()}
The $process()$ function acts as the main computation for the node. 
It modifies the local state of particles and variables, including creating new particles and variables and destroying old ones. 
It does not modify the system state.
\begin{align}
    &process : (\Omega_s \times Local) \rightarrow (\Omega_s \times Local)\\
    &process((c,\phi,\psi), (\phi_l,\psi_l)) = ((c,\phi,\psi), (\phi_l',\psi_l'))
\end{align}
This function is entirely application dependent.
When $c$ is a decision node, $c \in C_d$, the output $Local$ state shall contain information to determine the choice of the next node.

\subsubsection{push()}
The $push()$ function adds information from the local state to connected containers. 
This modifies the system state and preserves the local state.
\begin{align}
    &push : (\Omega_s \times Local) \rightarrow (\Omega_s \times Local)\\
    &push((c,\phi,\psi), (\phi_l,\psi_l)) = ((c,\phi',\psi'), (\phi_l,\psi_l))
\end{align}

The containers and environment nodes to which current node $c$ can push are the ones attached by push edges:
\begin{align}
    B_c = \{ b \in B \,|\, (c,b) \in E_{push}\} \\
    V_c = \{ v \in V \,|\, (c,v) \in E_{push}\}
\end{align}

The push function may only change containers connected by push edges, and then only to add a new object to the container\footnote{%
If an add is performed on a container to add a variable which already exists the behaviour is undefined and implementation dependent, updating is therefore done by pulling the variable to remove it and then push it to re-add it.
} with information from the local state:
\begin{align}
    \forall b \in B_c \,|\, \phi'(b)  ~\mathbf{\sf combinedwith}~ \phi(b) \\
    \forall b \in B \setminus B_c \,|\, \phi'(b)  = \phi(b) \\
    \forall v \in V_c \,|\, \psi'(v)  ~\mathbf{\sf combinedwith}~ \psi(v) \\
    \forall v \in V \setminus V_c \,|\, \psi'(v) = \psi(v) 
\end{align}


The default behaviour of push is to do nothing: $push(\omega, (\phi_l,\psi_l)) = (\omega, (\phi_l,\psi_l))$.

\subsubsection{next()}
The $next()$ function moves the control pointer to the next node and destroys the local state. 
This modifies the pointer node component of the system state.
\begin{align}
    &next : (\Omega_s \times Local) \rightarrow \Omega_s\\
    &next((c,\phi,\psi), (\phi_l,\psi_l)) = (c',\phi,\psi)
\end{align}

The nodes to which current node pointer $c$ can move to are defined by the control edge(s) from the current node:
\begin{align}
    c' \in \mbox{\bf \sf target} (c)
\end{align}

This set is a singleton set, except for decision nodes.
For decision nodes, the choice of which element to go to next is provided in the $Local$ information.

\end{document}